\documentclass[twocolumn]{aastex631}

\shorttitle{Carbon-bearing Molecules in a Possible Hycean Atmosphere}

\shortauthors{Madhusudhan et al.}

\graphicspath{{./}{figures/}}

\begin{document}

\title{Carbon-bearing Molecules in a Possible Hycean Atmosphere}

\author[0000-0002-4869-000X]{Nikku Madhusudhan}
\affiliation{Institute of Astronomy, University of Cambridge, Madingley Road, Cambridge CB3 0HA, UK}
\email{Correspondence: nmadhu@ast.cam.ac.uk}
\author{Subhajit Sarkar} \altaffiliation{These authors contributed comparably to this work.}\affiliation{School of Physics and Astronomy, Cardiff University, The Parade, Cardiff CF24 3AA, UK}

\author{Savvas Constantinou}\altaffiliation{These authors contributed comparably to this work.}\affiliation{Institute of Astronomy, University of Cambridge, Madingley Road, Cambridge CB3 0HA, UK}

\author{M\aa ns Holmberg}\altaffiliation{These authors contributed comparably to this work.}\affiliation{Institute of Astronomy, University of Cambridge, Madingley Road, Cambridge CB3 0HA, UK}

\author{Anjali A. A. Piette}
\affiliation{Earth \& Planets Laboratory, Carnegie Institution for Science, Washington, DC 20015, USA}

\author{Julianne I. Moses}
\affiliation{Space Science Institute, Boulder, CO 80301, USA}

\begin{abstract}
The search for habitable environments and biomarkers in exoplanetary atmospheres is the holy grail of exoplanet science. The detection of atmospheric signatures of habitable Earth-like exoplanets is challenging owing to their small planet-star size contrast and thin atmospheres with high mean molecular weight. Recently, a new class of habitable exoplanets, called Hycean worlds, has been proposed, defined as temperate ocean-covered worlds with H$_2$-rich atmospheres. Their large sizes and extended atmospheres, compared to rocky planets of the same mass, make Hycean worlds significantly more accessible to atmospheric spectroscopy with JWST. Here we report a transmission spectrum of the candidate Hycean world K2-18~b, observed with the JWST NIRISS and NIRSpec  instruments in the 0.9-5.2 $\mu$m range. The spectrum reveals strong detections of methane (CH$_4$) and carbon dioxide (CO$_2$) at 5$\sigma$ and 3$\sigma$ confidence, respectively, with high volume mixing ratios of $\sim$1\% each in a H$_2$-rich atmosphere. The abundant CH$_4$ and CO$_2$ along with the nondetection of ammonia (NH$_3$) are consistent with chemical predictions for an ocean under a temperate H$_2$-rich atmosphere on K2-18~b. The spectrum also suggests potential signs of dimethyl sulfide (DMS), which has been predicted to be an observable biomarker in Hycean worlds, motivating considerations of possible biological activity on the planet. The detection of CH$_4$ resolves the long-standing missing methane problem for temperate exoplanets and the degeneracy in the atmospheric composition of K2-18~b from previous observations. We discuss possible implications of the findings, open questions, and future observations to explore this new regime in the search for life elsewhere.
\end{abstract}


\section{Introduction} \label{sec:intro}

The detection and characterisation of habitable-zone exoplanets is a major frontier in modern astronomy. Until recently, the quest for exoplanetary habitability and biosignatures has been focused primarily on rocky exoplanets, naturally motivated by the terrestrial experience of life \citep{kasting1993,Meadows2018}. The extreme diversity of exoplanetary systems witnessed over the past three decades motivates considerations of new avenues in the search for life elsewhere. Such an endeavour may open the doors to a wider range of habitable environments that may be more numerous and more favourable to atmospheric characterisation. Hycean worlds, a recently proposed class of habitable exoplanets, represent one such avenue that is accessible to current observational facilities \citep{Madhusudhan2021}. 

Hycean worlds are a class of water-rich sub-Neptunes with planet-wide oceans underlying H$_2$-rich atmospheres. Such planets have a significantly wider habitable zone compared to terrestrial planets. With expected radii between 1-2.6 R$_\oplus$ for masses between 1-10 M$_\oplus$, Hycean planets represent a habitable subset of temperate sub-Neptunes that allow for a vast diversity of atmospheric and internal structures \citep{Madhusudhan2020,Madhusudhan2021,Piette2020, Nixon2021}. Such planets are also potentially abundant in the exoplanet population given the predominance of exoplanets in the sub-Neptune regime \citep{Fulton2018}. The large volatile content in the interior of a Hycean world implies a lower density and, hence, larger radius and lower gravity, compared to a rocky planet of comparable mass. The low gravity and low atmospheric mean molecular weight (MMW) in turn result in a larger atmospheric scale height for a given temperature relative to terrestrial-like exoplanets with high-MMW atmospheres. These factors make Hycean worlds readily accessible for atmospheric characterisation, including potential biomarker detection, using modest observing time with JWST \citep{Madhusudhan2021, Phillips2021, Phillips2022, Leung2022}.

The Hycean planet class was motivated by the demonstration that the bulk properties of the habitable-zone sub-Neptune K2-18~b \citep{Montet2015, cloutier2017, Cloutier2019, Benneke2019} are consistent with the possibility of a water-rich interior and a liquid-water ocean at habitable temperatures and pressures underlying a H$_2$-rich atmosphere \citep{Madhusudhan2020}. The planet has a mass of $8.63 \pm 1.35$ M$_\oplus$ and radius of $2.61 \pm 0.09$ R$_\oplus$, with an equilibrium temperature of $\sim$250-300 K for an albedo between 0-0.3 \citep{Cloutier2019,Benneke2019}. While a Hycean interpretation for K2-18~b is plausible and promising, a broad set of other internal structures and nonhabitable surface conditions are also compatible with its bulk properties \citep{Madhusudhan2020,Piette2020,Nixon2021}, especially when considering only cloud/haze-free atmospheres  \citep[e.g.][]{Scheucher2020,Pierrehumbert2023,Innes2023}. Originally, the observed transmission spectrum of the planet in the near-infrared (1.1-1.7 $\mu$m) with the Hubble Space Telescope (HST) WFC3 spectrograph suggested a H$_2$-rich atmosphere with strong H$_2$O absorption \citep{Benneke2019,Tsiaras2019,Madhusudhan2020}. However, other studies highlighted the degeneracy between H$_2$O and CH$_4$ in the observed HST spectrum \citep{Blain2021,Bezard2022}, and potential contributions due to stellar heterogeneities \citep{Barclay2021}, rendering the previous H$_2$O inference inconclusive. 

Atmospheric observations with JWST have the potential to provide important insights into the atmospheric, surface, and interior conditions of K2-18~b. The planet has been theoretically demonstrated to be accessible to detailed atmospheric characterisation with a modest amount of JWST time, including the possibility of detecting prominent CNO molecules, such as H$_2$O, CH$_4$, NH$_3$, as well as several biomarkers, such as (CH$_3$)$_2$S or dimethyl sulfide (DMS), methyl chloride (CH$_3$Cl), carbonyl sulfide (OCS), and others \citep{Madhusudhan2021}. The major molecules are expected to be detectable even in the presence of high-altitude clouds \citep{Constantinou2022}. Furthermore, several recent theoretical studies have demonstrated that atmospheric abundances of  prominent CNO molecules can be used to infer the presence of surfaces beneath H$_2$-rich atmospheres in temperate sub-Neptunes \citep{Yu2021,Hu2021,Tsai2021}. For example, the presence of an ocean underneath a shallow H$_2$-rich atmosphere, as would be the case for a Hycean world, may be inferred by an enhanced abundance of CO$_2$, H$_2$O and/or CH$_4$, but with a depletion of NH$_3$ \citep{Hu2021,Tsai2021,Madhusudhan2023}. 

In this work, we report the first JWST transmission spectrum of K2-18~b. The spectrum was observed using NIRISS SOSS and NIRSpec G395H instruments in the 0.9-5.2 $\mu$m wavelength range, which contains strong spectral features of multiple chemical species. The chemical constraints derived from the observed spectrum provide key insights into its atmospheric and surface conditions and pave the way for a new era of atmospheric characterisation of low-mass exoplanets with JWST. In what follows, we present our JWST observations and data reduction in section~\ref{sec:obs}. We discuss our atmospheric retrievals of the transmission spectrum in section~\ref{sec:retrieval}. We summarize our results and discuss the implications in section~\ref{sec:discussion}.

\section{Observations and Data Reduction} 
\label{sec:obs}

We report transmission spectroscopy of K2-18~b using the JWST NIRSpec \citep{ferruit2012, Birkmann2014} and NIRISS \citep{doyon_jwst_2012, Doyon2023} instruments. We observed two primary transits of the planet in front of its host star, one with each instrument, as part of the JWST GO Program 2722 (PI: N. Madhusudhan). The first transit was observed using the NIRSpec G395H grating between Jan 20, 2023, 18:37:38 UTC and Jan 21, 2023, 01:11:32 UTC for a total exposure time of 5.3 hours, which is nearly twice the expected transit duration. The observation was made in the Bright Object Time Series (BOTS) mode with the F290LP filter, the SUB2048 subarray and the NRSRAPID readout pattern, with the spectra dispersed over two detectors (NRS1 and NRS2). The two detectors, NRS1 and NRS2, span wavelength ranges of 2.73-3.72 $\mu$m and 3.82-5.17 $\mu$m, respectively, with a gap in between at 3.72-3.82 $\mu$m. The G395H grating offers the highest-resolution mode of NIRSpec with R$\sim$2700. The spectroscopic time-series observation is composed of 1625 integrations, with 12 groups per integration. For NIRSpec, the host star K2-18 was too bright for target acquisition (TA). Therefore, another nearby target (2MASSJ11301306+0735116) within the splitting distance of the science target was used for TA.

The second observation was conducted using the NIRISS Single Object Slitless Spectroscopy (SOSS) instrument mode \citep{Albert2023} between Jun 1, 2023, 13:49:20 UTC and Jun 1, 2023, 19:36:05 UTC, totalling an exposure time of 4.9 hours. The observation used the GR700XD grism (R$\sim$700), the CLEAR filter, the SUBSTRIP256 subarray and the NISRAPID readout pattern, giving a wavelength coverage of 0.85– 2.85 $\mu$m for the first spectral order. The exposure consisted of 648 integrations, with 4 groups per integration. There were no tilt events or high-gain antenna movements during any of the observations.

\begin{figure*}
	\includegraphics[width=0.5\textwidth]{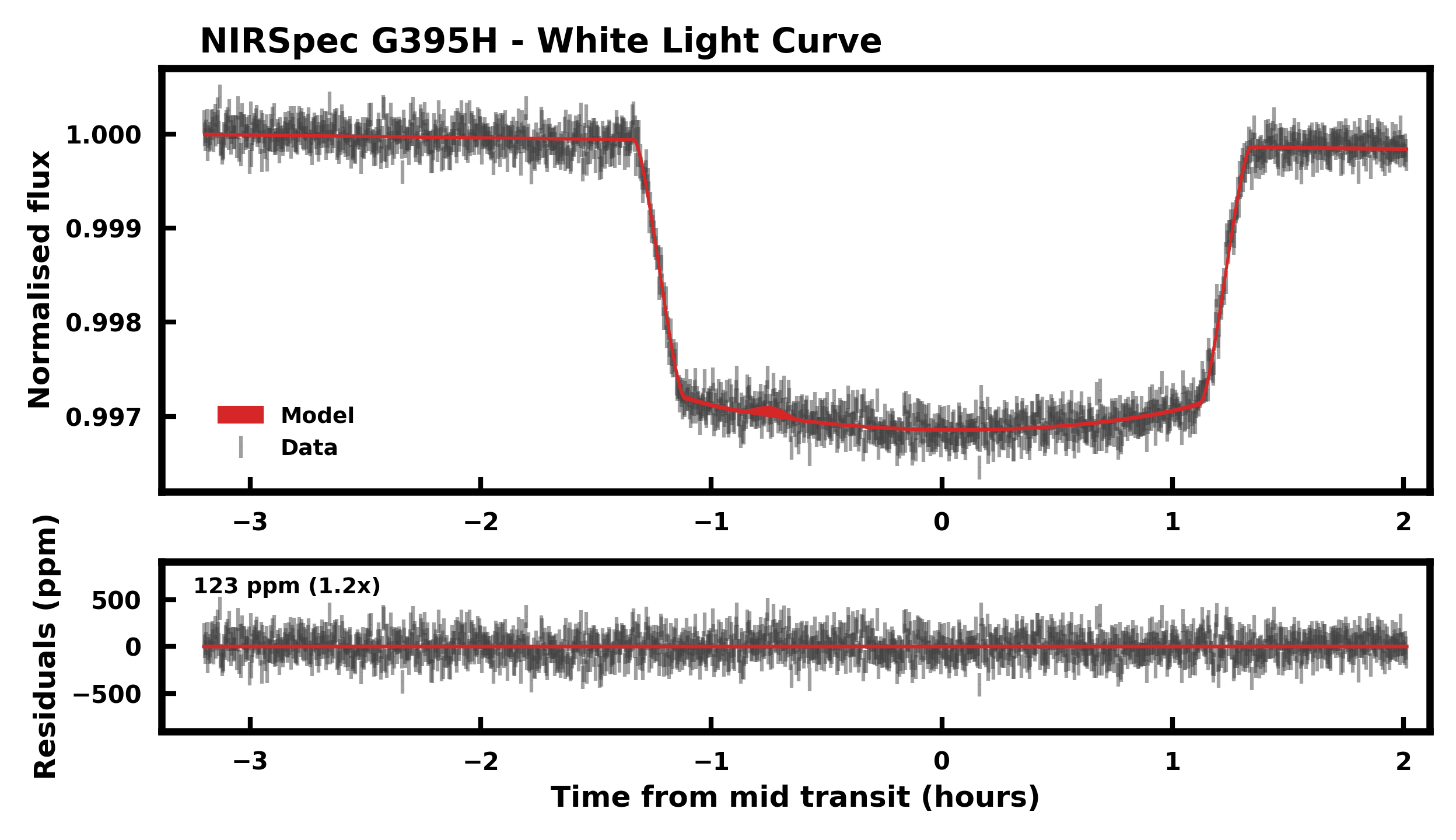}
    \includegraphics[width=0.5\textwidth]{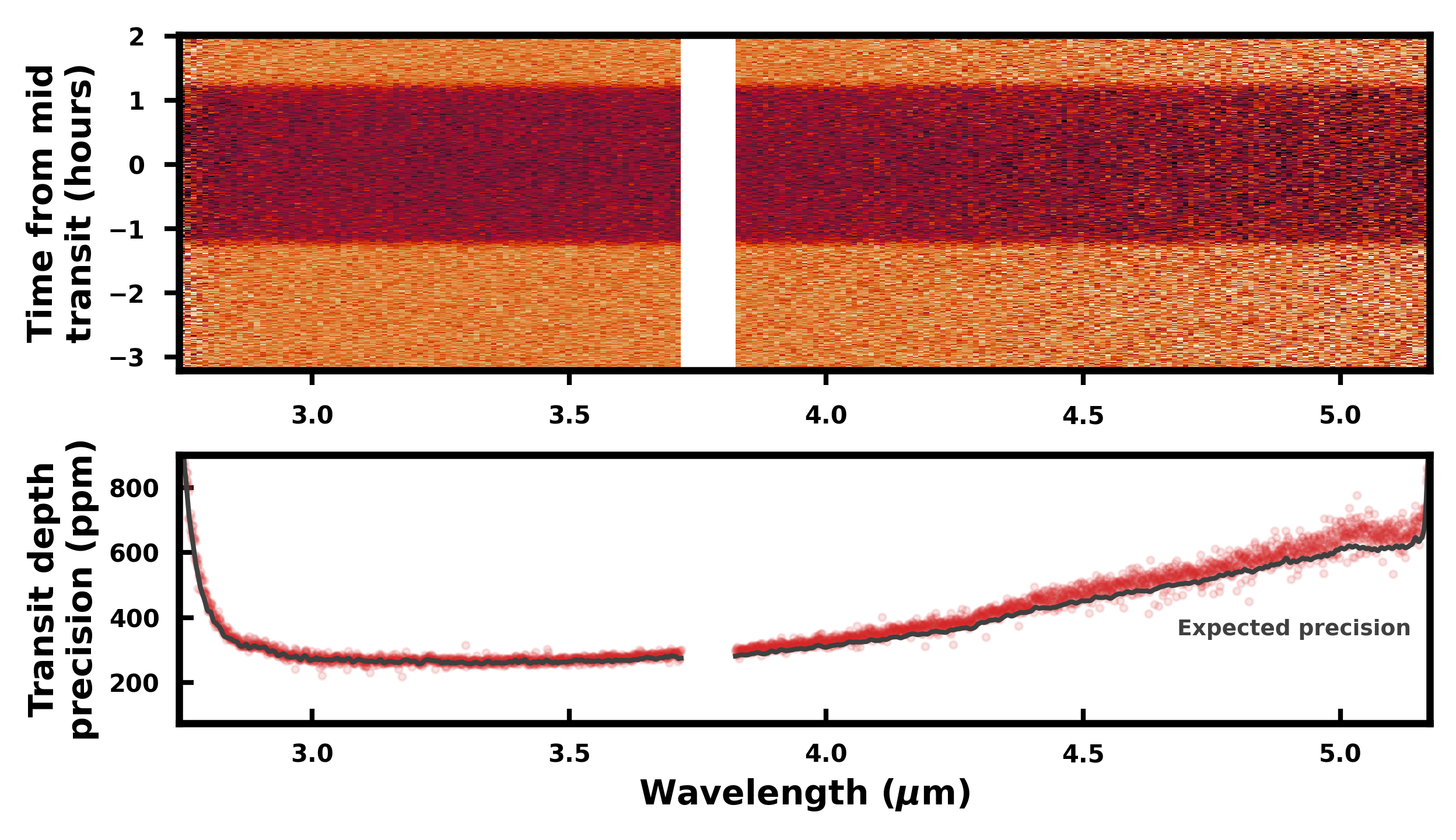}
    \caption{Left: white light curve for the transit of K2-18 b observed with NIRSpec G395H. The top panel shows the combined white light curve from NRS1 and NRS2, together with the 1$\sigma$ model interval in red. The bottom panel shows the residuals after subtracting the median model, together with line at zero. The standard deviation of the residuals is measured to be 123 ppm, which is 1.2$\times$ the expected noise level. The red line is shown to indicate zero. Right: the top panel shows the normalised spectroscopic light curves (binned in wavelength for visual clarity). The detector gap is shown in white. The bottom panel depicts the transit depth precision at pixel resolution.
    }
    \label{fig:wlc_nirspec} 
\end{figure*}

\begin{figure*}
	\includegraphics[width=0.5\textwidth]{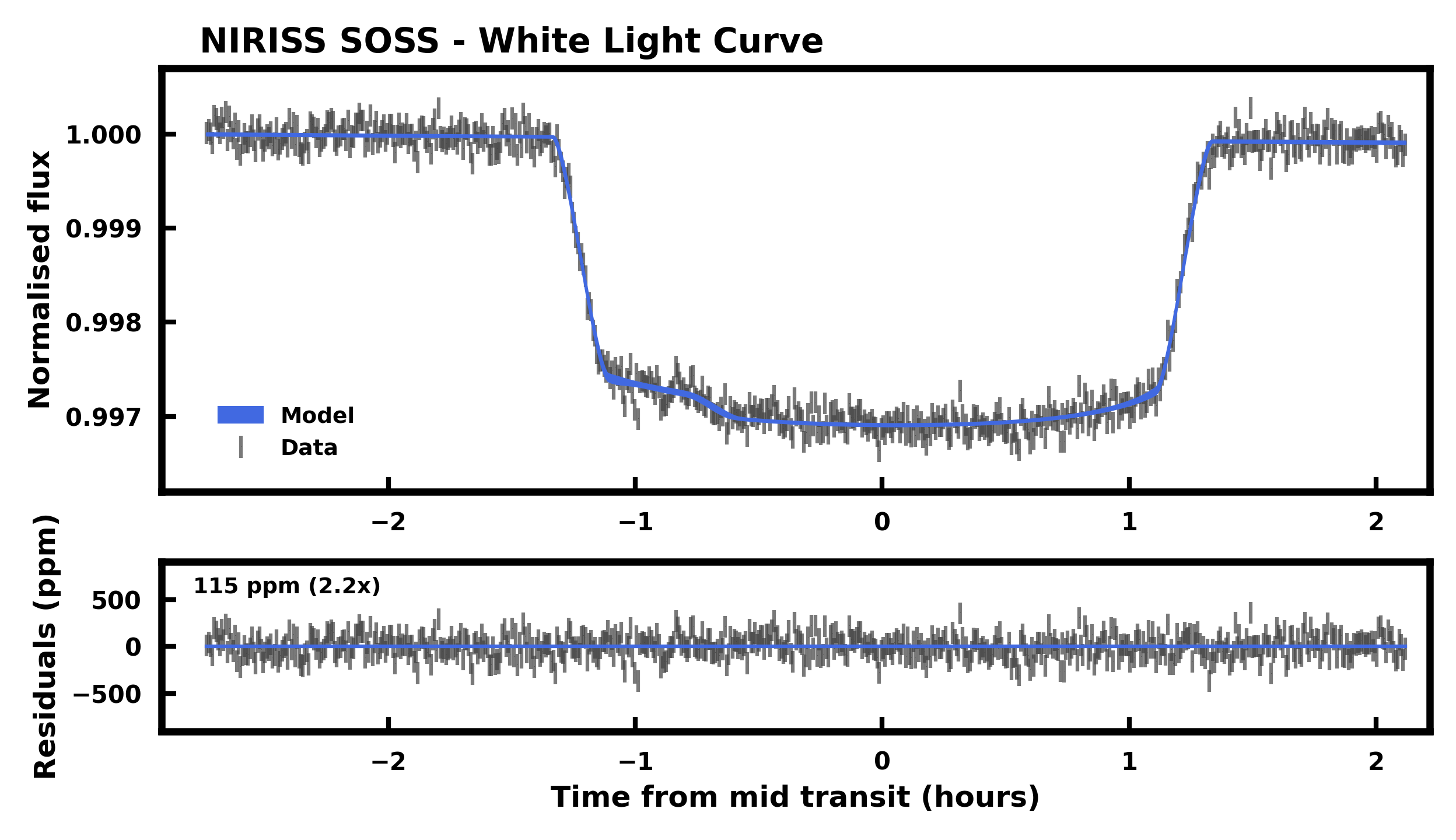}
    \includegraphics[width=0.5\textwidth]{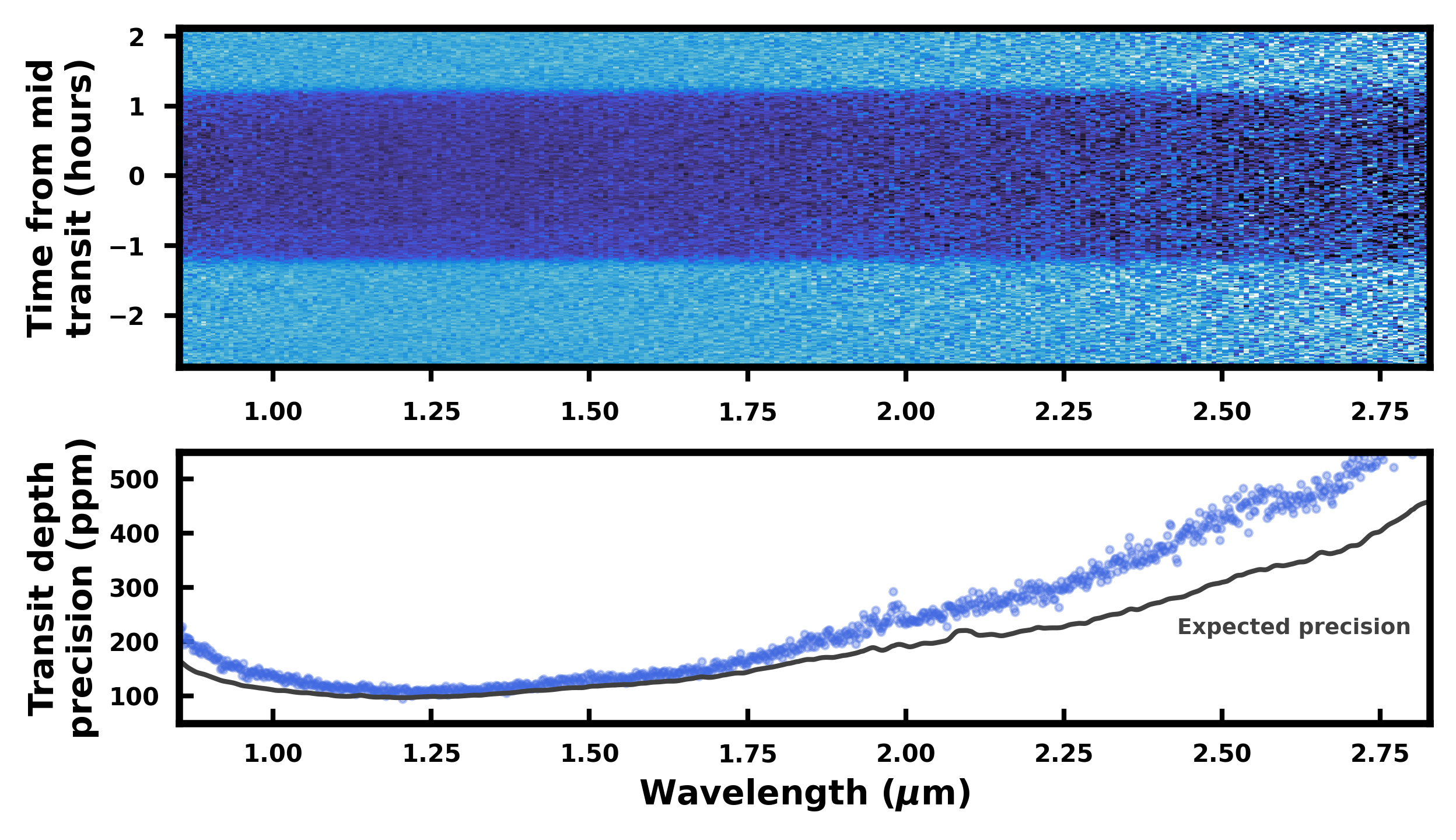}
    \caption{Left: white light curve for the transit of K2-18 b observed with NIRISS SOSS. The blue contour shows the 1$\sigma$ model interval. We find evidence of a spot occultation at the start of the transit. The bottom panel shows the residuals after subtracting the median model. The standard deviation of the residuals is measured to be 115 ppm, which is 2.2$\times$ the expected noise level.  The blue line is shown to indicate zero. Right: the top panel shows the normalised spectroscopic light curves (binned in wavelength for visual clarity). The bottom panel shows the transit depth precision obtained from the light-curve fitting, binned at two pixel columns per light curve.
    }
    \label{fig:wlc_niriss} 
\end{figure*}

\subsection{NIRSpec}

The data reduction is conducted using a combination of the JWST Science Calibration Pipeline \citep{Bushouse2020} and our custom-built pipeline for the spectral extraction. We start with the raw 2D images (the .uncal files) of the spectroscopic time-series data which contain group-level counts for each integration. Stage 1 of the data reduction is performed mainly using the JWST Science Calibration Pipeline. This involves performing saturation flagging, superbias subtraction, reference pixel correction, linearity correction, dark current subtraction (where reference data is available), jump detection and linear fitting of the group-level ramps to obtain the count rate for each pixel per integration. This is repeated for each integration in the exposure. The jump detection threshold is set at 5$\sigma$.  Prior to ramp fitting, we also perform an additional step for background subtraction, at the group level, in order to mitigate 1/f noise, as common in previous works \citep{JWST_ERS2023, Rustamkulov2023, Alderson2023}. For this step we measure the background level for each detector column, as the mean of the pixels that are $\pm$10 pixels away from the midpoint of the curved trace, while masking bad pixels and cosmic-ray hits. The outputs of Stage 1 are 2D images of the count rate for each integration, saved as .rateints files. 

We also use steps from the JWST Science Calibration Pipeline for Stage 2, which applies the wavelength calibration for the spectral trace. Following previous studies using NIRSpec \citep[e.g.][]{Alderson2023}, we forgo the flat-field correction in Stage 2 for our differential transit measurement. The resultant 2D images along with the wavelength calibration, which are saved as .calints files, are then used for the spectral extraction of the time series of 1D stellar spectra.

We conduct the spectral extraction, Stage 3, applying our custom-built pipeline to convert the 2D images into 1D spectra. This is conducted for each of the two detectors (NRS1 and NRS2) separately. We first create a bad pixel mask based on the data quality flags in the Stage 2 products. We then extract the 1D spectrum from the 2D image using an optimal extraction algorithm \citep{horne_optimal_1986}. To perform the extraction, we obtain the point-spread functions (PSFs) as the sum of the first three principal components of the time series of the detector images\footnote{For this the sum of the detector columns was normalised to unity and bad pixels were interpolated.}, inspired by the principal-component-analysis-based morphology analysis in \cite{coulombe_broadband_2023}. This takes into account the wavelength and time dependence of the PSF. Outliers were iteratively rejected during the spectrum extraction, with the threshold set at 5$\sigma$. Spectral channels with more than 20\% of the flux masked were discarded from further analysis.

\subsection{NIRISS}

To reduce the NIRISS data, we used the JWST Science Calibration Pipeline \citep{Bushouse2020} for Stages 1-2 and the JExoRES pipeline for the spectral extraction \citep{holmberg2023}. During Stage 1, we perform the standard saturation flagging, superbias subtraction, linearity correction, jump detection (threshold set at 5$\sigma$) and fitting of the group-level ramp to obtain the count rate. We perform a custom background subtraction step before the linearity correction and ramp fitting to reduce the effect of 1/f noise. In line with \cite{Radica2023} and \cite{Albert2023}, this involves temporarily subtracting a model of the flux of the detector, containing both the background and stellar flux, to reveal the 1/f noise. Initially, we use the groupwise median stack of all integrations to model the flux. For each group and integration, the 1/f noise level is then estimated using the median of each column, while masking the traces of the spectral orders, as well as bad pixels. This 1/f noise estimate is subtracted from the data to reduce the level of correlated noise. We later repeat all stages of the NIRISS data reduction using updated models of the background level and stellar flux to refine the 1/f noise correction step, as described further below. Furthermore, we do not perform the dark current subtraction, given the quality of the dark reference file from the Calibration Reference Data System (CRDS) is insufficient, so as not to contribute additional noise \citep{Feinstein2023, Radica2023}.

For Stage 2, we perform the flat-field correction before modelling the background flux in a two-step process. First, we use the background model available via JDOX, created from program 1541, scaled to the median of all integrations in a small rectangular region ($x \in [720, 770]$, $y \in [210, 250]$). By subtracting this we precisely remove the brighter of the two background components, caused by zodiacal light. However, this leaves residuals in the dimmer component of the background for columns up to the $\sim$700 pixel column.  This is corrected by first generating a median image from all integrations, and then from this median image obtaining the median of each pixel column (while masking spectral traces, contamination and bad pixels). These column-wise medians are then subtracted from each integration image. This is only performed for columns up to the 700-pixel column. We use this additional background flux to update the (scaled) background model. Next, we perform the order tracing, PSF estimation, and spectral extraction (Stage 3) according to \cite{holmberg2023}, with an extraction aperture of 35 pixels, leaving out the background refinement since we correct the 1/f noise at the group level. As in the case of NIRSpec, spectral channels with more than 20\% of the flux masked were discarded from further analysis.

Finally, we repeat the data reduction stages in order to improve the 1/f noise correction. This time, we model the flux at the group level using the updated background model from above and the groupwise out-of-transit median stellar flux. The background model is scaled using the integration time of each group while the out-of-transit stellar flux (minus the background) is scaled with a model of the transit light curve, derived using the initial white light curve from the first spectral order. We note that in the remaining analysis we only consider the first spectral order given that the second order has a considerably lower flux level, meaning that it is more sensitive to systematics due to sources of contamination.

\begin{figure*}
	\includegraphics[width=\textwidth]{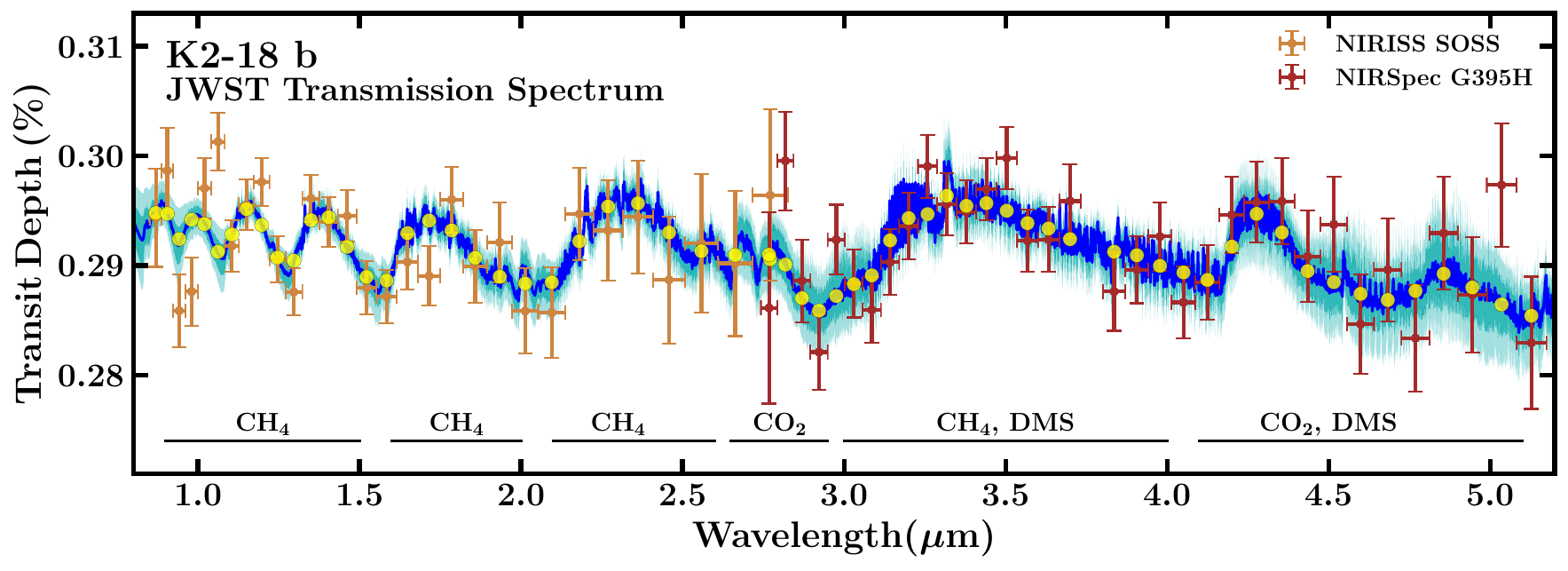}
    \caption{The transmission spectrum of K2-18~b. The observed JWST spectrum and retrieved model fits are shown for the one-offset retrieval case discussed in section~\ref{sec:retrieval}. The data in orange show our NIRISS spectrum between 0.9 - 2.8 $\mu$m and those in dark-red show our NIRSpec G395H spectrum between 2.8 - 5.2 $\mu$m. The spectra are binned to $R \approx 25$ and $R  \approx  55$ for NIRISS and NIRSpec, respectively, for visual clarity. The retrievals are conducted on the native resolution spectra, resulting in the best-fit reduced $\chi_\nu^2 = 1.080$. The NIRSpec spectrum is vertically offset by $-41$ ppm, corresponding to the median retrieved offset in the one-offset retrieval case. The blue curve shows the median retrieved model spectrum, while the medium- and lighter-blue contours denote the 1$\sigma$ and 2$\sigma$ intervals, respectively. Yellow points correspond to the median spectrum binned to match the observations. The prominent molecules responsible for the features in different spectral regions are labelled.
    }
    \label{fig:spectrum} 
\end{figure*}

\subsection{Starspot Occultations} \label{sec:spot}

Starspot and faculae crossings during exoplanet transits are known to affect the apparent transit depth \citep[e.g.][]{Pont2008, Czesla2009}. If left uncorrected, starspot and faculae occultations effectively decrease and increase the observed transit depth, respectively. This effect is wavelength dependent and must therefore be accounted for so as not to impact the transmission spectrum. From the present observations we find evidence of starspot occultations, which is especially strong for the transit observed with NIRISS. Fortunately, correcting the transmission spectrum is possible by measuring the intensity ratio and size of the occulted feature from the transit light curves.

We perform a joint inference of the transit parameters and the properties of the active region using the semianalytical spot modelling code \texttt{SPOTROD} \citep{spotrod2014}. Hence, we also avoid having to discard the affected data. The code computes the transit light curve with arbitrary limb darkening (affecting the stellar photosphere and spots equally) and homogeneous circular star spots or faculae. The spot is represented by four parameters: the spot-to-star radius ratio $R_\mathrm{spot}/R_*$, the spot-to-unspotted stellar surface intensity ratio $f$, and the coordinates of the spot centre projected onto the stellar surface ($\vartheta$, $r^2$). Intensity ratios below and above 1 represent starspots and faculae, respectively. Similar to \cite{Espinoza2019}, we use nested sampling \citep{Skilling2004}, implemented via \texttt{MultiNest} \citep{Feroz2009}, to obtain the Bayesian model evidence and parameter estimation. This allows us to perform a Bayesian model comparison between transit models with or without stellar spots. For this, we use uniform priors between 0 and 0.3 for the spot-to-star radius ratio, 0 and 2 for the intensity ratio, 0 and $2\pi$ for $\vartheta$ (with periodic boundary condition), and 0 and 1 for $r^2$, ensuring uniform sampling of the disk.

Using the NIRISS white light curve from the first order, we find that a single starspot is strongly preferred over no spots with a Bayes factor of $\ln B = 21.6$, corresponding to a significance of $6.9\sigma$. Fig. \ref{fig:wlc_niriss} shows the white light curve and the fitted model. We obtain an intensity ratio of $f = 0.9329_{-0.0091}^{+0.0082}$ and a spot-to-star radius ratio of $R_\mathrm{spot}/R_* = 0.254_{-0.044}^{+0.033}$. For the NIRSpec white light curve (NRS1 and NRS2 combined), including a spot is also marginally preferred by the data compared to a model without spots, with a Bayes factor of $\ln B = 1.15$, corresponding to a significance of $2.1\sigma$. The spot parameters in this case are $f = 0.82_{-0.39}^{+0.60}$ and $R_\mathrm{spot}/R_* = 0.113_{-0.077}^{+0.094}$. For these reasons, we choose to use the spot modelling for both observations.

\begin{table*}
\centering
\begin{tabular}{lcccc}
\hline \hline
Parameter & NIRISS & NIRSpec & Weighted Average \\ \hline
Midtransit time, T$_0$ (BJD - 2400000.5) & $60096.729368_{-0.000065}^{+0.000063}$ &  $59964.969453_{-0.000034}^{+0.000035}$ &  - \\ 
Inclination, $i$ ($^{\circ}$) & $89.550_{-0.020}^{+0.021}$ & $89.567_{-0.011}^{+0.012}$  & $89.563 \pm 0.010$ \\ 
Normalised semi-major axis, $a / R_*$ & $79.9_{-1.4}^{+1.4}$ & $80.92_{-0.72}^{+0.78}$ & $80.68 \pm 0.68$ \\ 
Planet-to-star radius ratio, $R_\mathrm{p} / R_*$ & $0.05412_{-0.00017}^{+0.00019}$ & $0.05441_{-0.00021}^{+0.00018}$ & - \\ 
First LDC, $u_1$ & $0.065_{-0.047}^{+0.080}$ & $0.291_{-0.079}^{+0.046}$ & - \\ 
Second LDC, $u_2$ & $0.220_{-0.105}^{+0.066}$ & $-0.098_{-0.055}^{+0.097}$ & - \\ \hline
\end{tabular}
\caption{Parameter estimation from the white light curve analysis of our JWST NIRISS SOSS and NIRSpec G395H observations of K2-18 b. For the white light curve of NIRSpec, we combine NRS1 and NRS2. For both white light curves, we use a linear trend to model the baseline flux as well as the spot modelling described in section \ref{sec:spot}. The orbital period is held fixed at $P = 32.940045$ days, adopted from \protect\cite{Benneke2019}.
}
\label{tab:wlc_params}
\end{table*}

\subsection{Light-curve Analysis} \label{sec:light_curves}

The 1D spectral time series from both observations are then used for light-curve fitting to derive the transit depths. This is done in three stages. The first stage uses the white light curves, from both observations, to derive the wavelength-independent system parameters at high precision. For NIRISS we use the white light curve from the first order and for NIRSpec we use the combined white light curve from both NRS1 and NRS2. In the next stage, we bin the light curves to $R \sim 20$ and fit the wavelength-dependent limb-darkening coefficients (LDCs). Finally, we fix these LDCs in the respective $R \sim 20$ bins and fit the transit depths at native resolution to obtain the final transmission spectrum of the planet.

As described above, we model the transit light curves using \texttt{SPOTROD} \citep{spotrod2014} as we find evidence of starspot occultations. We assume a circular orbit with the orbital period from \cite{Benneke2019}. We adopt the two-parameter quadratic limb-darkening law, in line with previous JWST transmission spectroscopy studies of M dwarf systems \citep[e.g.][]{moran_high_2023, lustig-yaeger_jwst_2023} and previous work on K2-18 b \citep{Benneke2019}. For the LDCs, we use the parameterization and priors by \cite{Kipping2013}. To model the baseline flux we implement both a linear and a quadratic trend. We find that the NIRISS out-of-transit white light curve shows a preference for a linear trend, while the NIRSpec data show a weak preference for a quadratic trend in the case of NRS1 and a moderate preference in the case of NRS2. For the white light curves, we use a linear trend for both instruments, given that we find the derived system parameters to be insensitive to the choice of trend. However, for the spectrum, we consider both trends for NIRSpec. For NIRISS, we fix the trend to be linear. Moreover, we discard the first 5 minutes of both observations owing to a small settling ramp. Apart from starspot occultations, no other systematics can be identified, highlighting the excellent data quality provided by JWST.

For the white light curve of each observation (see left panels of Figs.~\ref{fig:wlc_nirspec} and \ref{fig:wlc_niriss}), we fit for the midtransit time (T$_{0}$,), the normalised semi-major axis ($a/R_*$), the orbital inclination ($i$), the planet-to-star radius ratio ($R_\mathrm{p} / R_*$), the quadratic LDCs ($u_1$ and $u_2$), two parameters representing the baseline flux, and four spot parameters (described in \ref{sec:spot}). In the likelihood, we also include a parameter to inflate the photometric uncertainties to match the residual scatter between the data and the transit light-curve model. We measure the precision to be 1.2$\times$ and 2.2$\times$ the expected noise level (photon and read noise, propagated using the Jacobian of the transit model) for the white light curves from NIRSpec (NRS1 and NRS2 combined) and NIRISS, respectively. To sample the posteriors and estimate the parameters we use \texttt{MultiNest} \citep{Feroz2009}, resulting in the parameters given in Table \ref{tab:wlc_params}. For all cases, we find that the derived system parameters are consistent within $1\sigma$, regardless of the trend and the choice to model the starspot or not.

Next, we bin the spectroscopic light curves to $R \sim 20$ for the purpose of fitting the wavelength-dependent LDCs. This resolution strikes a balance between precision and the expected wavelength variability of the limb darkening. We choose to fit the LDCs instead of using values from stellar atmospheric models in order to maximise accuracy \citep{Csizmadia2013, espinoza_limb_2015}. We fit these binned light curves using MultiNest and fix the system parameters (T$_{0}$, $a/R_*$, $i$) to the values obtained from the white light curve analysis (see Table \ref{tab:wlc_params}), i.e. the weighted average $a/R_*$, $i$, and T$_{0}$ from each observation. We also fixed $R_\mathrm{spot}/R_*$, $\vartheta$, and $r^2$ to the best-fit parameters for each white light curve. Equipped with empirical LDCs, we go on to fit the transit depths of the high-resolution light curves, while fixing the LDCs to the values within their respective $R \sim 20$ bin. This leaves only $R_\mathrm{p} / R_*$, $f$, the 2-3 trend parameters, and the uncertainty scaling parameter as free parameters. We fit the light curves at the pixel level for NIRSpec and two pixels per bin for NIRISS, given the potential inaccuracy of the NIRISS SOSS wavelength calibration \citep{Albert2023}. To fit these high-resolution light curves we use the Levenberg–Marquardt algorithm, as applied in previous works \citep{Alderson2023, moran_high_2023}. The right panel of Figs. \ref{fig:wlc_nirspec} and \ref{fig:wlc_niriss} show the precision of the resulting transmission spectrum at high resolution. In total, we have 1010 and 3401 spectral data points for NIRISS and NIRSpec, respectively, covering the 0.9 - 5.2 $\mu$m wavelength range.

As mentioned above, we obtain two spectra in the case of NIRSpec, one with a linear trend and another with a quadratic trend. For the quadratic-trend case, we construct white light curves for NRS1 and NRS2, and fit these separately to obtain detector-averaged values for the quadratic trend component\footnote{We model the baseline flux as $F(t) = F_{\textrm{out}} (1 + p_1 t + p_2 t^2)$, where $F_{\textrm{out}}$ is the out-of-transit flux at the start of the observation, $p_1$ is the linear-trend parameter, and $p_2$ is the quadratic-trend parameter.}. For each detector, we then fix the quadratic trend component to these values when fitting the spectroscopic light curves \citep{moran_high_2023}. Overall, we find that the NRS1 spectrum is almost agnostic to the choice of trend ($\sim$10 ppm difference), whereas, for NRS2, there is a significant difference between the two spectra (approximately a 60 ppm offset). Since we only have one transit observation for NIRSpec, we use an offset as a free parameter in the atmospheric retrieval to account for potential baseline shifts.

\section{Atmospheric Retrieval} \label{sec:retrieval}

The observed transmission spectrum allows us to retrieve the atmospheric properties of K2-18~b at the day-night terminator region. We perform the retrieval using the AURA retrieval code \citep{Pinhas2018} following a similar approach to previous retrieval studies of the planet considering HST and/or simulated JWST observations \citep[e.g.][]{Madhusudhan2020,Madhusudhan2021,Welbanks2019,Constantinou2022}. The planet's terminator is modelled as a plane-parallel atmosphere in hydrostatic equilibrium, with uniform chemical composition. The chemical abundances and pressure-temperature ($P$-$T$) profile are free parameters in the model. The retrieval framework follows a free chemistry approach, whereby the individual mixing ratio of each chemical species is a free parameter. The atmospheric temperature structure is modelled with a parametric $P$-$T$ profile \citep{Madhusudhan2009} with six free parameters. The model also considers inhomogeneous clouds/hazes at the day-night terminator region \citep{macdonald2017,Pinhas2018}. They are modelled as a combination of a grey cloud deck at a parametric cloud-top pressure $P_{c}$, above which are hazes with Rayleigh-like spectral contributions, with an enhancement factor $a$ and a scattering slope $\gamma$; for H$_2$ Rayleigh scattering, $a$ = 1 and $\gamma$ = -4. The combined clouds and hazes cover a fraction $\phi$ of the atmosphere at the terminator region.

The model includes molecular opacity contributions from prominent CNO molecules expected in temperate H$_2$-rich atmospheres as considered in previous works \citep{Pinhas2018,Welbanks2019,Constantinou2022}. These include H$_2$O, CH$_4$, NH$_3$, HCN, CO and CO$_2$. The molecular absorption cross sections are obtained following recent works \citep{gandhi2017,gandhi2020} using line lists from the following sources: H$_2$O \citep{Polyansky2018}, CH$_4$ \citep{Hargreaves2020}, NH$_3$ \citep{Coles2019}, HCN \citep{harris2006, barber2014}, CO \citep{rothman2010, Li2015} and CO$_2$ \citep{HUANG2013, huang2017}. Pressure broadening due to H$_2$ is considered for all these molecules as described in \citet{gandhi2020}.

We additionally consider five molecules that have been suggested to be promising biomarkers in habitable rocky exoplanets \citep{Segura2005, domagal-goldman2011, seager2013a, seager2013b, catling2018, schwieterman2018} as well as Hycean worlds \citep{Madhusudhan2021}: (CH$_3$)$_2$S (or DMS), CS$_2$, CH$_3$Cl, OCS and N$_2$O. The absorption cross sections of CH$_3$Cl, OCS and N$_2$O were computed from the corresponding line lists from the \textsc{HITRAN} database \citep{HITRAN2016}: CH$_3$Cl \citep{ch3cl_1, ch3cl_2}, OCS \citep{ocs_1, ocs_2, ocs_3, ocs_4, ocs_5, ocs_6, ocs_7}, and N$_2$O \citep{n2o_2}.  As pressure broadening due to H$_2$ is not available for these molecules, we only use thermal broadening. For DMS and CS$_2$, we use the absorption cross sections provided directly by \textsc{HITRAN} \citep{dms_cs2_1,dms_cs2_2,HITRAN2016} at 1 bar and 298 K, following \cite{Madhusudhan2021}. In addition to molecular cross-sections, we also consider H$_2$-H$_2$ and H$_2$-He collision-induced absorption \citep{borysow1988,orton2007,abel2011,richard2012}.

\begin{figure*}
	\includegraphics[width=\textwidth]{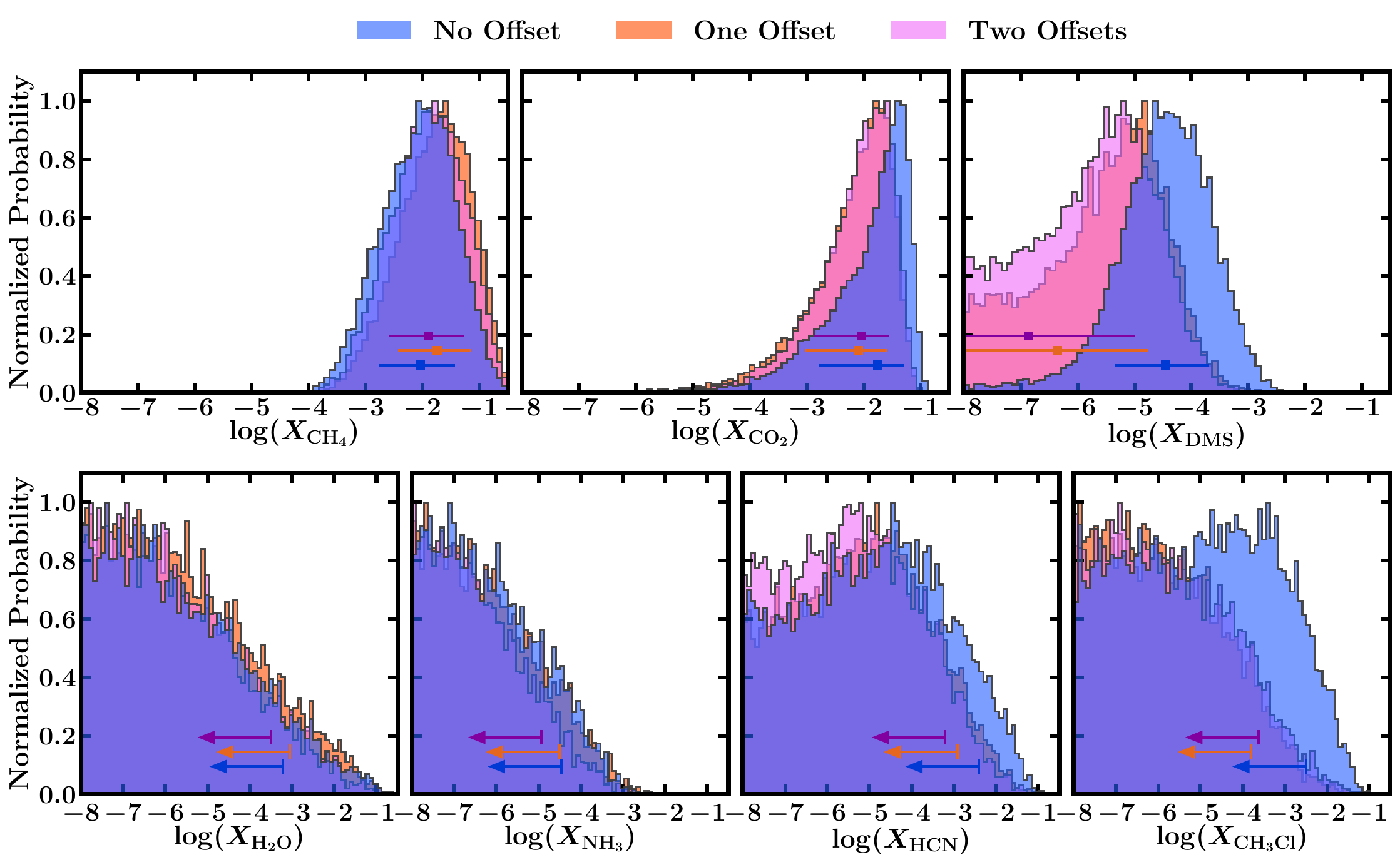}
 \caption{Retrieved posterior probability distributions for the mixing ratios of important molecules for the three retrieval cases described in section \ref{sec:retrieval}. The horizontal error bar in the top row denotes each distribution's median and corresponding 1$\sigma$ interval for the three molecules with significant detections. The arrows in the bottom row indicate 95\% upper limits. We find strong evidence for CH$_4$ and CO$_2$, at a significance of 5$\sigma$ and 3$\sigma$, respectively. We find marginal evidence for DMS and no significant evidence for the remaining molecules. The abundance estimates and detection significances are shown in Table~\ref{tab:abundances}.
 }
\label{fig:posteriors}
\end{figure*}

\begin{table*}
\centering
\begin{tabular}{lcccccccc}
\hline \hline
Cases & CH$_4$ & CO$_2$ & DMS & H$_2$O & NH$_3$ & CH$_3$Cl & CO & HCN  \\
\hline
No offset & $-2.04_{-0.72}^{+0.61}$ (4.7$\sigma$) & $-1.75_{-1.03}^{+0.45}$ (2.9$\sigma$) & $-4.46_{-0.88}^{+0.77}$ (2.4$\sigma$) & \textless$-3.21$ & \textless$-4.46$ & \textless$-2.50$ & \textless$-3.00$ & \textless$-2.41$ \\ 
1 offset & $-1.74_{-0.69}^{+0.59}$ (5.0$ \sigma$) & $-2.09_{-0.94}^{+0.51}$ (2.9$\sigma$) & $-6.35_{-3.60}^{+1.59}$ ($\sim$1$\sigma$) & \textless$-3.06$ & \textless$-4.51$ & \textless$-3.80$ & \textless$-3.50$ & \textless$-2.92$\\ 
 2 offsets & $-1.89_{-0.70}^{+0.63}$ (5.0$\sigma$) & $-2.05_{-0.84}^{+0.50}$ (3.2$\sigma$) & $-6.87_{-3.25}^{+1.87}$ ($-$) & \textless$-3.49$ & \textless$-4.93$ & \textless$-3.62$ & \textless$-3.19$ & \textless$-3.21$ \\ \hline
\end{tabular}
\caption{Retrieved molecular abundances and detection significances of prominent molecules in the atmosphere of K2-18~b.
The three canonical atmospheric model cases are described in section~\ref{sec:retrieval}, and pertain to different considerations for offsets between data from different instruments. The molecular abundances are shown as $\log_{10}$ of volume mixing ratios. The retrieved median and 1$\sigma$ estimates are given for CH$_4$, CO$_2$ and DMS which show strong to marginal detections, and 95\% upper-limits are given for the remaining molecules. The quantities in brackets for CH$_4$, CO$_2$ and DMS show the detection significances greater than 1$\sigma$. The detection significances have a nominal statistical uncertainty of $\sim$0.1$\sigma$ due to the uncertainty on the Bayesian evidence estimated by the nested sampling algorithm. See section \ref{sec:detection_sigs}.
}
\label{tab:abundances}
\end{table*}

\subsection{Retrieval Setup}
\label{sec:retrieval_setup}
Our canonical model comprises of 22 free parameters overall: 11 corresponding to the individual mixing ratios of the above chemical species, 6 for the $P$-$T$ profile, 4 for the clouds/hazes and 1 for the reference pressure $P_\mathrm{ref}$, defined as the pressure at a fixed planetary radius of 2.61 R$_\oplus$. The Bayesian inference and parameter estimation is conducted using the \texttt{MultiNest} nested sampling algorithm \citep{Feroz2009} implemented through \texttt{PyMultiNest} \citep{Buchner2014}. The retrieval setup and priors on the model parameters are similar to those in recent implementations of the AURA retrieval 
framework \citep{Madhusudhan2020, Constantinou2022} and are shown in Appendix~\ref{sec:priors}. We also consider variations to the canonical model, including a cloud/haze-free atmosphere, Mie scattering due to hazes, and the presence of stellar heterogeneities influencing the spectrum, as discussed below. We use the present JWST NIRISS and NIRSpec transmission spectra of K2-18 b in the 0.9 - 2.8 $\mu$m and 2.8 - 5.2 $\mu$m ranges, respectively, at their native resolution for the retrieval. As discussed in section \ref{sec:light_curves}, we derive two spectra for NIRSpec corresponding to the two trends (linear and quadratic) we use to model the baseline flux; the NIRISS spectrum has a robust preference for a linear trend. Furthermore, the NIRSpec G395H grating uses two detectors (NRS1 and NRS2) and recent studies have suggested the possibility of an offset in the spectrum derived from a given detector \citep{moran_high_2023}. Therefore, we conducted retrievals on different combinations of NIRSpec spectra obtained with  different trends and offsets.

We consider two broad combinations of data: (1) NIRISS and linear-trend NIRSpec spectra, and (2) NIRISS and quadratic-trend NIRSpec spectra. For each combination, we consider a range of different spectral offsets as free parameters in the retrieval. We consider four cases: a baseline case with no offsets, one combined offset for the NIRSpec spectrum, one offset for the NIRISS spectrum, and two separate offsets for the NIRSpec NRS1 and NRS2 spectra. The last case is the most conservative and is motivated by the transit depth offset between NRS1 and NRS2 recently reported by \cite{moran_high_2023}. The offset on either NIRISS or NIRSpec represents cases where we assume no offset between NIRSpec NRS1 and NRS2 \citep{Alderson2023, Lustig2023, August2023} but instead allow for an offset between NIRISS and NIRSpec as a whole. In the no-offset case, we consider the data as is and do not perform any offsets. We conduct the four retrievals for each data combination and assess their relative Bayesian evidence.

\begin{table*}
\centering
\begin{tabular}{lccccccccc}
\hline \hline
Cases & $T_{10 \mathrm{mbar}}$ (K) & $\phi$ & $\log a$ & $\gamma$ & $\log (P_{\mathrm{c}}\mathrm{/bar})$ & $\log (P_{\mathrm{ref}}\mathrm{/bar})$ & OS1/ppm & OS2/ppm & $\chi_\nu^2$ \\
\hline 
No offset & $257_{-74}^{+127}$ & $0.63_{-0.21}^{+0.22}$ & $7.31_{-2.64}^{+1.76}$ & $-11.67_{-3.17}^{+3.65}$ & $-0.55_{-1.20}^{+0.99}$ & $-1.88_{-0.53}^{+0.48}$ & - & - & 1.082\\
1 offset & $242_{-57}^{+79}$ & $0.63_{-0.15}^{+0.19}$ & $8.20_{-1.89}^{+1.24}$ & $-11.11_{-2.61}^{+2.94}$ & $-0.51_{-1.20}^{+0.98}$ & $-2.48_{-0.47}^{+0.47}$ & $-41_{-13}^{+11}$ & - & 1.080\\ 
2 offsets & $235_{-56}^{+78}$ & $0.64_{-0.16}^{+0.19}$ & $8.21_{-1.92}^{+1.25}$ & $-11.34_{-2.62}^{+2.93}$ & $-0.46_{-1.15}^{+0.95}$ & $-2.34_{-0.50}^{+0.47}$ & $-30_{-13}^{+12}$ & $26_{-16}^{+15}$ & 1.081\\ \hline
\end{tabular}
\caption{Retrieved temperature, cloud/haze properties and reference pressure for the atmosphere of K2-18~b, as well as retrieved dataset offsets where applicable. Similarly to Table~\ref{tab:abundances}, the three canonical atmospheric model cases are described in section~\ref{sec:retrieval}, which pertain to different considerations for offsets (denoted OS above) between data from different instruments. The 1-offset retrieval considers a shift to the NIRSpec observations relative to NIRISS, while the 2-offset retrieval considers two shifts, applied to observations from the NIRSpec NRS1 and NRS2 detectors. The temperature constraints shown are at 10~mbar, which corresponds to the observed photosphere. In all cases, the best-fit reduced $\chi_\nu^2$ is close to unity, with the degrees of freedom being 4389, 4388, and 4387 for no, 1, and 2 offsets, respectively.
}
\label{tab:physical_properties}
\end{table*}

By comparing the Bayesian evidence of the considered cases, we find that generally some offset is preferred over no offsets. For the linear-trend data, a single offset between the data sets (on either NIRISS or NIRSpec) is the most preferred. The two one-offset cases are comparable, with an offset on NIRSpec being marginally favoured over that on NIRISS. For the quadratic-trend data, the two-offset case is the most favoured. The fact that the configuration with separate offsets for NRS1 and NRS2 is strongly preferred for the quadratic trend while not being favoured for the linear trend suggests that, for the present observation, using a quadratic trend contributes to an offset between NRS1 and NRS2. Based on these findings, from across the different combinations we finally select three nominal cases for NIRSpec along with NIRISS: (a) NIRSpec with a linear trend and no offsets, representing the data without modification, (b) NIRSpec with a linear trend and one offset, and (c) NIRSpec with a quadratic trend and two separate offsets for NRS1 and NRS2, representing the most conservative case. These are the retrieval cases considered in the rest of this work. In what follows we report the atmospheric properties at the day-night terminator region of K2-18~b retrieved using the transmission spectrum in each of these cases, as shown in Tables~\ref{tab:abundances} and \ref{tab:physical_properties}.

\subsection{Prominent CNO Molecules}
Our atmospheric retrieval provides important constraints on the dominant CNO molecules expected in H$_2$-rich atmospheres. The retrieved spectral fit is shown in Fig.~\ref{fig:spectrum}, and the corresponding posterior distributions for several molecules are shown in Fig.~\ref{fig:posteriors}. Amongst the prominent CNO molecules, we find strong spectral contributions from CH$_4$ and  CO$_2$ in a H$_2$-rich atmosphere. For our retrieval with no offset, we derive log volume mixing ratios of $\log$(X$_{\text{CH}_4}$) = $-2.04^{+0.61}_{-0.72}$ and $\log$(X$_{\text{CO}_2}$) = $-1.75^{+0.45}_{-1.03}$. For the one-offset case, we obtain $\log$(X$_{\text{CH}_4}$) = $-1.74^{+0.59}_{-0.69}$ and $\log$(X$_{\text{CO}_2}$) = $-2.09^{+0.51}_{-0.94}$. As shown in Table~\ref{tab:abundances}, these abundance estimates for both molecules are consistent across the three retrieval cases, with median abundances of $\sim$1\% and average uncertainties below 1-dex, underscoring the robustness of the derived estimates. Both CH$_4$ and CO$_2$ are detected for the first time in a sub-Neptune exoplanet and the precision of their abundance estimates is the best measured for any molecule in a sub-Neptune atmosphere to date.

We do not find significant contributions due to H$_2$O or NH$_3$, but find 95\% upper limits of -3.21 for $\log$(X$_{\text{H}_2\text{O}}$) and -4.46 for $\log$(X$_{\text{NH}_3})$ in the no-offset case. These upper limits are also consistent with those from the other retrieval cases, as shown in Table~\ref{tab:abundances}. The nondetections of both molecules are important considering their strong spectral features and detectability expected in the 0.9-5.2 $\mu$m range \citep{Madhusudhan2021, Constantinou2022}. The nondetection of H$_2$O is at odds with its previous inference using the HST WFC3 spectrum in the 1.1-1.7 $\mu$m range \citep{Tsiaras2019,Benneke2019,Madhusudhan2020}. A strong degeneracy between H$_2$O and CH$_4$ in the HST WFC3 band was noted previously \citep{Blain2021, Bezard2022}. Our retrieved CH$_4$ abundance is consistent with previous predictions of stronger absorption due to CH$_4$ relative to H$_2$O in the HST WFC3 band \citep{Blain2021, Bezard2022} and some upper bounds on the CH$_4$ abundance \citep{Madhusudhan2021,Blain2021,Bezard2022}. Our results, therefore, resolve the degeneracy in the atmospheric composition of K2-18 b from previous observations.

While our NIRISS spectrum is generally in good agreement with the previous HST WFC3 spectrum \citep{Benneke2019} in the 1.1-1.7 $\mu$m range, as shown in Fig.~\ref{fig:wfc3_comparison}, there is notable difference in two of the data points at the blue end of the WFC3 spectrum. A comparison between the new JWST spectrum and the HST spectrum is presented in Appendix \ref{sec:hst_comparison}. Furthermore, the presence of multiple CH$_4$ features across our NIRISS and NIRSpec spectral range provides a very strong detection of CH$_4$, as discussed in section~\ref{sec:detection_sigs}. We note that our upper limit for H$_2$O corresponds to the planet's stratosphere at pressures below $\sim$100 mbar. Water vapor may very well be abundant at deeper levels in the atmosphere, but condensation of H$_2$O is expected in the upper troposphere of this temperate planet \citep{Benneke2019,Madhusudhan2023}, resulting in a comparatively dry stratosphere, as on Earth.

We also do not detect CO or HCN despite their strong spectral features expected in the 0.9-5.2 $\mu$m range \citep{Madhusudhan2023}. The 95\% upper limits for both molecules are shown in Table~\ref{tab:abundances}, with maximum values of -3.00 for $\log$(X$_{\text{CO}}$) and -2.41 for $\log$(X$_{\text{HCN}})$. 
Given the low-temperature H$_2$-rich atmosphere, the nondetection of CO is not necessarily surprising, as CH$_4$ is expected to be the dominant equilibrium constituent in deep H$_2$ atmospheres on cooler planets \citep{Moses2013}. However, some CO is expected to be present from disequilibrium quenching in deep atmospheres or photochemistry at high altitudes, becoming especially important in thinner atmospheres \citep{Yu2021,Tsai2021,Hu2021,Madhusudhan2023}. The high abundances of CO$_2$ and CH$_4$, along with the nondetection of NH$_3$ and CO, and a high CO$_2$/CO ratio, are consistent with predictions for an ocean surface under a thin H$_2$-rich atmosphere \citep{Hu2021,Madhusudhan2023}, as discussed further in section~\ref{sec:discussion}.

\vspace{5mm}
\subsection{Biosignature Molecules}

The retrievals provide notable constraints on two methyl-group terrestrial biomarkers, DMS and CH$_3$Cl, predicted to be detectable in Hycean atmospheres, especially for K2-18~b \citep{Madhusudhan2021}. We retrieve a DMS mixing ratio of $\log$(X$_{\text{DMS}}$) = $-4.46^{+0.77}_{-0.88}$ in the no-offset case, $-6.35^{+1.59}_{-3.60}$ in the one-offset case, and $-6.87^{+1.87}_{-3.25}$ in the two-offset case. The weaker constraints on the DMS abundance with increasing number of offsets between the detectors are due to the DMS spectral feature being broad and the continuum level spanning multiple detectors, as discussed in section~\ref{sec:detection_sigs}. The potential inference of DMS is of high importance as it is known to be a robust biomarker on Earth and has been extensively advocated to be a promising biomarker for exoplanets \citep{Seager2013,seager2016,catling2018,Madhusudhan2021}; this is discussed further in section \ref{sec:discussion}. 
We also find a nominal peak in the posterior distribution of CH$_3$Cl that is more significant than other nondetections such as those of H$_2$O or NH$_3$, as shown in Fig.~\ref{fig:posteriors}. The retrieval in the no-offset case provides an abundance estimate of $\log$(X$_{\text{CH$_3$Cl}}$) = $-6.62^{+3.08}_{-3.40}$ and a 95\% upper limit of -2.50, which are comparable within $\sim$1-dex to those from the other retrieval cases.  

We note that CH$_4$, DMS, and CH$_3$Cl all have strong spectral features owing to the C-H bond in the 3-3.5$\mu$m range in the NIRSpec G395H band and are therefore degenerate to some extent, as shown in Fig.~\ref{fig:contributions}. However, thanks to the multiple strong CH$_4$ features in the NIRISS band, the degeneracies between the two stronger molecules, CH$_4$ and DMS, are somewhat mitigated, whereas CH$_3$Cl is relatively unconstrained.

\subsection{Molecular Detection Significances} \label{sec:detection_sigs}
In addition to the abundance constraints discussed above, we determine the detection significances of the key molecules using Bayesian model comparisons \citep{Trotta2008, Benneke2013, Pinhas2018}. In the present context, we evaluate the detection significance of a molecule as a Bayesian preference for a model fit to the data while including that molecule, relative to the same model with the molecule absent \citep{ Benneke2013, Pinhas2018}. Naturally, such an evidence comparison depends to some extent on the specific model considerations and the combination of the data used, which are discussed 
in section~\ref{sec:retrieval_setup}. Therefore, we estimate detection significances for the prominent molecules for each of the three retrieval cases, as shown in Table~\ref{tab:abundances}. We note that the detection significances reported here have intrinsic statistical uncertainties of $\sim$0.1$\sigma$ owing to the uncertainty in Bayesian evidence obtained by the nested sampling algorithm for a given retrieval.

Among the prominent CNO molecules, we find the strongest detection for CH$_4$ at 4.7-5.0$\sigma$ across all three cases. The consistently high detection significance value independent of the offset(s) considered underscores the robustness of the detection, which is due to multiple features of CH$_4$ being present across the 1-5 $\mu$m range of the observed spectrum as shown in Fig.~\ref{fig:spectrum}. We also detect CO$_2$ robustly at $\sim$3$\sigma$ significance across all three cases. The strong detection of CO$_2$ is made possible by its prominent spectral feature around 4.3 $\mu$m and the full feature along with the spectral baseline around it being on the same detector (NRS2) in the NIRSpec band. As discussed above, we do not find significant evidence for NH$_3$, H$_2$O, CO or HCN. 

Among the biomarkers, we find some evidence for DMS depending on the retrieval case. The detection significance of DMS depends on the offsets considered. This is because, while DMS has a strong spectral feature around 3.3 $\mu$m in the NIRSpec NRS1 detector, the feature is broad and the spectral baseline falls on neighbouring detectors --- the NIRISS at shorter wavelengths and NIRSpec NRS2 at longer wavelengths. Therefore, the spectral amplitude and hence the detection significance and abundance estimate rely strongly on the relative offsets between the detectors; the detection significance lowers for each additional offset in the retrieval. We infer DMS at 2.4$\sigma$ confidence for the no-offset case but at only $\sim$1$\sigma$ for the one-offset case and no significant evidence for the two-offset case. Nevertheless, as shown in Fig.~\ref{fig:posteriors}, in all three cases the retrieved posterior distributions for DMS show notable peaks within 1-dex of each other, except that for the cases with offsets the distributions contain long low-abundance tails owing to the degeneracy with the spectral baseline as discussed above. The posteriors are also notably  different from the nondetections for other prominent molecules, such as H$_2$O, NH$_3$ or HCN, and indicate a nonnegligible chance for DMS being present in the atmosphere. Upcoming observations will be able to further constrain the presence of DMS, as discussed below and in section~\ref{sec:discussion}.

\begin{figure*}
	\includegraphics[width=\textwidth]{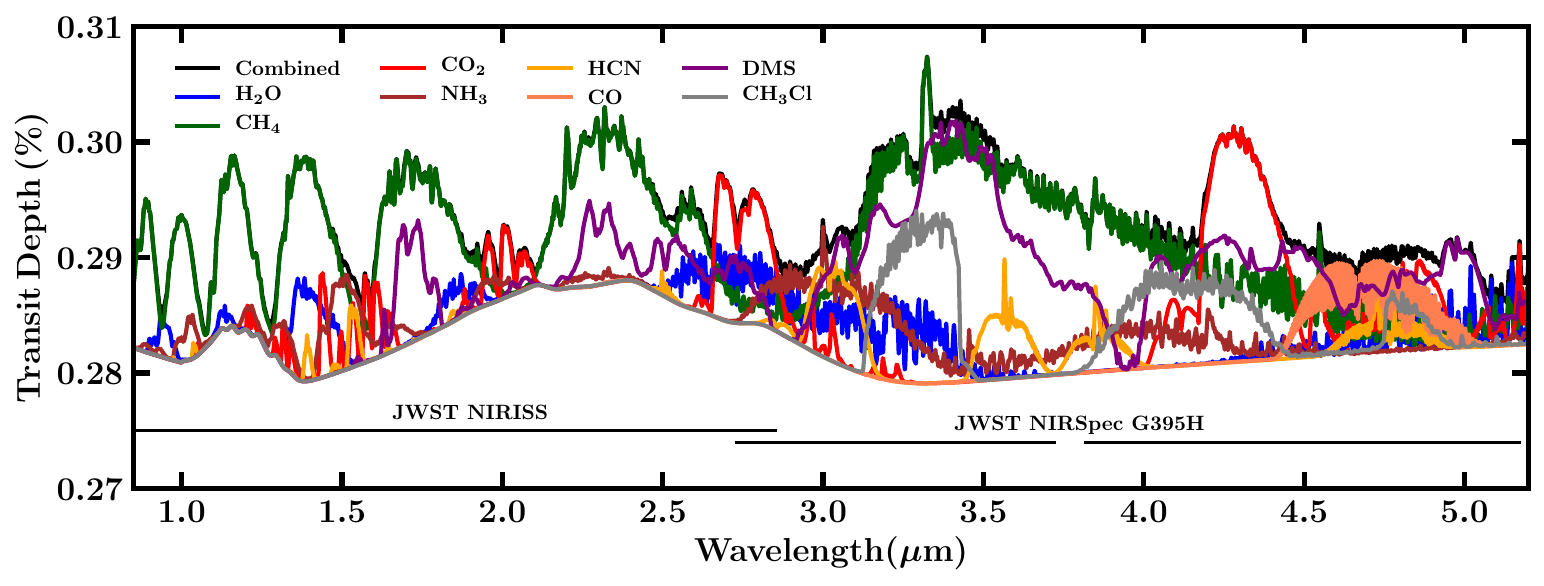}
    \caption{Spectral contributions of key molecular species in the 1-5 $\mu$m range. The different curves show individual contributions from different molecules to a nominal model transmission spectrum of K2-18~b shown in black and denoted as "Combined". The model assumes a mixing ratio of 10$^{-2}$ for CH$_4$ and CO$_2$, 10$^{-4}$ for H$_2$O, and 10$^{-5}$ for all the other species, consistent with our retrieval estimates discussed in section~\ref{sec:retrieval}, and an isothermal temperature profile of 250 K. Each curve corresponds to a transmission spectrum with opacity contributions from a single molecule at a time, in addition to H$_2$-H$_2$ and H$_2$-He collision-induced absorption. The spectral ranges of our JWST NIRISS and NIRSpec observations are also indicated; the NIRSpec range spans two detectors (NRS1 and NRS2), with a gap between them at 3.72-3.82 $\mu$m.}
    \label{fig:contributions} 
\end{figure*}

There could also be potential contributions from CH$_3$Cl to the spectrum, albeit without any appreciable detection significance, as evident from Fig.~\ref{fig:posteriors}.  
CH$_3$Cl, being a methylated molecule like DMS, has some overlapping features with DMS as shown in Fig.~\ref{fig:contributions}. Therefore, its significance increases marginally in the absence of DMS. While we do not detect CH$_3$Cl on its own in any of the retrieval cases, the combination of DMS and CH$_3$Cl has a slightly higher detection significance of 2.7$\sigma$ than DMS alone (2.4$\sigma$) in the no-offset case. We find no significant evidence for any other biomarkers considered in the retrievals.

Overall, we find CH$_4$ and CO$_2$ to be our most confident detections, followed by DMS, with the abundance estimates reported above. While our results provide important first insights into the chemical composition of K2-18~b, upcoming observations will be able to verify our present findings. These include observations of the transmission spectrum of K2-18~b with JWST MIRI between $\sim$5 and 10 $\mu$m (JWST Program GO 2722; PI: N. Madhusudhan) and more observations with JWST NIRSpec G395H and G235H (JWST Program GO 2372, PI: R. Hu). 

\subsection{Clouds/Hazes and Photospheric Temperature} \label{sec:clouds}

The observed transmission spectrum provides nominal constraints on the presence of clouds/hazes in the atmosphere. The constraints on the cloud/haze  parameters are shown in Table~\ref{tab:physical_properties}. The constraints on the cloud-top pressure for the gray clouds are relatively weak, mostly lying below the observable photosphere (e.g. cloud-top pressures $\gtrsim$100 mbar). Even though the scattering slope ($\gamma$) is not well constrained, the enhancement factors ($a$) are generally higher than that expected for H$_2$ Rayleigh scattering ($a$=1), albeit still consistent with the latter at the 3$\sigma$ uncertainties for the no-offset case. The haze coverage fraction at the day-night terminator region is constrained to $\sim$0.6, albeit with large uncertainties of $\sim$0.2. Based on Bayesian model comparisons as discussed above, we find that a model with clouds/hazes is preferred over a model without clouds/hazes by 2.8-3.2$\sigma$ across the three retrieval cases considered. However, more observations in the optical to near-infrared wavelengths would be needed to provide stronger constraints on the cloud/haze properties of K2-18 b, as discussed in Appendix \ref{sec:no_clouds}. We find that the abundance constraints on the molecules are consistent within the 1-$\sigma$ uncertainties between the retrievals with and without clouds/hazes. 

We additionally consider retrievals in which the parametric clouds/hazes of the canonical model are replaced with Mie scattering hazes, as described by \citet{Pinhas2017} and \citet{Constantinou2023}. We specifically include two forms of organic haze, using optical constants presented by \citet{Khare1984} and \citet{He2023}. We find no evidence for this model, and a Bayesian model comparison shows a preference in favour of the parametric clouds/hazes considered in our canonical model. While the haze properties are unconstrained in this retrieval, the abundance constraints on the gaseous species remain consistent with those from our canonical retrieval cases.

The observations provide nominal constraints on the temperature in the planetary photosphere. We find the temperature at 10 mbar to range between 235$^{+78}_{-56}$ and 257$^{+127}_{-74}$ among the three cases, as shown in Table \ref{tab:physical_properties}. We note that the retrieved temperature structure is typically less well constrained using transmission spectroscopy, compared to emission spectroscopy \citep{Madhu2016}. Nevertheless, the retrieved temperature range and the nondetection of H$_2$O allow for the possibility of H$_2$O clouds in the deeper atmosphere. Considering the pressures probed by the spectral features across our observed range, we find that the photosphere, i.e. the  $\tau$=1 surface, lies between pressures of $\sim$0.1-100 mbar. The 10 mbar temperature estimates for the three retrieval cases are shown in Table~\ref{tab:physical_properties}. While the H$_2$-rich atmosphere can result in a significant greenhouse effect and warm the ocean surface, clouds and/or hazes play a crucial role in cooling the atmosphere and decreasing the temperature gradient \citep{Madhusudhan2020, Madhusudhan2021, Madhusudhan2023, Piette2020}. The possible presence of clouds can allow more temperate conditions at the ocean surface compared to those predicted by cloud-free models \citep[see][]{Innes2023}.

\subsection{Stellar Heterogeneities}
\label{sec:stellar_heterogeneities}
We also consider retrievals including the effects of unocculted stellar heterogeneities on the transmission spectrum, using our AURA retrieval framework \citep{Pinhas2018}. We do not find significant evidence for such effects in any of the three retrieval cases. The spot covering fraction in our retrieval is consistent with zero at the 2$\sigma$ uncertainties, and a model with stellar heterogeneity is not favored over a model without it across the three cases. We also find that the abundance constraints on the molecules are not significantly affected by the consideration of stellar heterogeneities in the retrieval. The nondetection of H$_2$O is further evidence against the effects of unocculted stellar heterogeneities on the present transmission spectrum, given that H$_2$O in cool unocculted starspots may contaminate the spectrum, as recently reported \citep{Barclay2021,moran_high_2023}. 

\section{Summary and Discussion} \label{sec:discussion}

We report a transmission spectrum of the candidate Hycean exoplanet K2-18~b observed with JWST. The spectrum observed with the JWST NIRISS and NIRSpec instruments spans the 0.9-5.2 $\mu$m range containing strong absorption features of prominent CNO molecules and biomarkers predicted for Hycean worlds. We report strong detections of CH$_4$ and CO$_2$ in a H$_2$-rich atmosphere at 5$\sigma$ and 3$\sigma$ confidence, respectively, with high volume mixing ratios ($\sim$1\%) for both. However, we do not detect H$_2$O, NH$_3$, CO or HCN, while obtaining upper limits on their abundances that are consistent with chemical expectations for an ocean under a cold and thin H$_2$-rich atmosphere \citep{Hu2021,Madhusudhan2023}. We also find potential evidence for DMS, which has been predicted as a robust biomarker in both terrestrial and Hycean worlds. These findings support the Hycean nature of K2-18 b and the potential for biological activity on the planet.

The observed mass, radius, and equilibrium temperature of K2-18 b have been known to be consistent with a degenerate set of internal structures \citep{Madhusudhan2020,Madhusudhan2023}. These include (a) a Hycean world with a thin H$_2$-rich atmosphere over a water-rich interior, (b) a mini-Neptune with a deep H$_2$-rich atmosphere, or (c) a predominantly rocky super-Earth interior with a deep H$_2$-rich atmosphere. Our retrieved chemical composition of the atmosphere of K2-18 b helps distinguish between these scenarios. In what follows, we discuss the implications and possible explanations of our findings and future directions.

\subsection{A Potential Hycean World}
K2-18~b has been originally predicted to be the archetype of a Hycean world \citep{Madhusudhan2021}, one with habitable oceans underneath a H$_2$-rich atmosphere. The currently derived chemical composition of the atmosphere is in agreement with previous theoretical predictions for the presence of an ocean under a shallow H$_2$-rich atmosphere \citep{Hu2021, Madhusudhan2023}. In particular, \cite{Hu2021} predicted the abundance of CO$_2$ to range between 4$\times$10$^{-4}$ and 10$^{-1}$ and that of CH$_4$ to range between 1.5-5.3 $\times$10$^{-2}$, which are in agreement with our retrieved abundances. They also predicted relatively lower abundances of CO, NH$_3$ and stratospheric H$_2$O, which are consistent with our nondetections. A low retrieved H$_2$O gas-phase mixing ratio at pressures less than $\sim$100 mbar is consistent with condensation due to a tropospheric cold trap \citep{Madhusudhan2023}, as in Earth's stratosphere, and with the retrieved thermal structure in this work and previous studies \citep[e.g.][]{Benneke2019,Piette2020,  Madhusudhan2023}. That is, H$_2$O could be abundant below its condensation region in the atmosphere, but the transit observations of the terminator region do not probe deep enough to detect it.

The main argument against K2-18~b being a potential Hycean world is based on climate considerations --- a greenhouse effect in a thick H$_2$-rich atmosphere that is cloud/haze-free would result in temperatures being sufficiently elevated at pressures greater than $\sim$10 bar, such that a liquid ocean would instead be converted to a steam-dominated atmosphere that ultimately goes supercritical at depth for irradiation levels relevant to K2-18~b \citep{Scheucher2020, Piette2020, Innes2023, Pierrehumbert2023}. Any global ocean surface must then reside at pressures less than $\sim$10 bar. However, too shallow a H$_2$ atmosphere could be subject to escape over time \citep[e.g.,][]{Kubyshkina2018a, Kubyshkina2018b, Hu2023}, so there is a limited parameter range over which a habitable ocean could exist on K2-18~b without significant clouds/hazes. As mentioned previously, high-albedo tropospheric water clouds or scattering hazes can help alleviate the steam and supercritical water problem by reducing the stellar energy absorbed by the planet \citep{Madhusudhan2021, Piette2020}.

Our retrieved atmospheric temperatures and upper limits for H$_2$O are consistent with the possibility that H$_2$O is condensing into clouds below the photosphere in K2-18~b, indicating a cold upper troposphere \citep[see also][]{Benneke2019, Madhusudhan2023}. Furthermore, as discussed in section~\ref{sec:clouds}, our retrievals show some evidence for scattering due to hazes at the day-night terminator region of the atmosphere; however, more optical observations are required to robustly confirm the same. If such clouds or hazes enshroud the planet or, in particular, are present on the dayside, they could provide the required albedo to sustain a habitable ocean in K2-18~b. Given the possible greenhouse effects of clouds themselves, a relatively shallow H$_2$ atmosphere might still be required to maintain low-enough temperatures for a liquid ocean.

\subsection{Is a Deep Atmosphere a Possibility?}

Our derived atmospheric composition in combination with K2-18~b's irradiation level is seemingly inconsistent with a deep-atmosphere, mini-Neptune scenario. In that scenario any photochemically produced carbon and nitrogen species would be recycled in the deep atmosphere back to their thermodynamically stable forms, CH$_4$ and NH$_3$, with transport then returning these molecules back to the observable upper regions of the atmosphere \citep{Yu2021,Tsai2021,Madhusudhan2023}. Although CO$_2$ can have a substantial mixing ratio in a H$_2$-rich atmosphere with a high metallicity and/or low C/O ratio \citep[e.g.,][]{Moses2013}, we find that the simultaneous presence of $\sim$1\% CO$_2$ and CH$_4$ in a thick H$_2$-rich atmosphere requires a moderately high C/H metallicity, a very low C/O ratio, ($\sim$0.02) and efficient vertical quenching. For example, C/H = 30$\times$ solar, O/H = 690$\times$ solar, and $K_{zz}$ $\gtrsim$ 10$^7$ cm$^2$ s$^{-1}$ combined with a temperature profile with $T_{int}$ = 60 K provides CH$_4$ and CO$_2$ mixing ratios of the right magnitude.  Aside from the question of how such an oxygen-enriched atmosphere would originate from a formation and evolution standpoint, this scenario would cause H$_2$O to dominate over H$_2$ throughout much of the atmosphere below a few hundred mbar (increasing its mean molecular weight), and NH$_3$ and CO would also be present in nonnegligible quantities, none of which are consistent with the observations.

Nor can our observations be explained by a shallow, H$_2$-rich atmosphere overlying a solid surface at the few-bar pressure level, despite the abundances of CH$_4$ and CO$_2$ being seemingly consistent with such models  \citep{Yu2021,Tsai2021,Madhusudhan2023}. The planet's bulk density precludes a thin H$_2$ atmosphere overlying an extensive silicate mantle \citep{Madhusudhan2020}. Even a pure silicate interior would require a thick ($\gtrsim$10$^3$ bar) H$_2$-rich envelope to explain the mass and radius. Although interaction of a thick H$_2$ atmosphere with a deep silicate mantle might explain the presence of some atmospheric CO$_2$ \citep[e.g.,][]{Kite2020, Kite_Barnett2020, Schlichting2022, Tian2023}, such models do not generally predict a resulting 1\% CO$_2$ mixing ratio from high-pressure interaction with silicates at depth in a thick H$_2$ atmosphere, nor can that scenario explain the apparent lack of recycling of N$_2$ back to NH$_3$. Overall, the bulk density of the planet, combined with our derived chemical composition for the atmosphere, presents strong evidence in support of K2-18~b as a Hycean world rather than a rocky or volatile-rich planet with a deep H$_2$ atmosphere, or a rocky planet with a thin H$_2$ atmosphere. 

One caveat to the above discussion is that current photochemical models for mini-Neptune conditions assume ideal-gas behavior and do not consider how a primordial H$_2$ atmosphere with its expected reduced species might interact chemically with a supercritical water layer and/or silicate magma at depth in the atmosphere. Moreover, our understanding of the bulk composition and chemistry of super-Earth and mini-Neptune atmospheres is rudimentary at this stage, and the depleted NH$_3$ and CO could potentially have some chemical explanation that has not been considered up to this point.
We also note that the retrieved abundances represent average chemical abundances in the observable photosphere ($\sim$0.1-100 mbar) at the day-night terminator region. While the molecules with robust detections, CH$_4$ and CO$_2$, have been predicted to be relatively uniform in this pressure range for K2-18~b in several cases \citep{Hu2021,Yu2021,Madhusudhan2023} more precise observations in the future may be able to constrain nonuniform chemical abundances in the atmosphere.

\subsection{Possible Evidence of Life}
\label{sec:possible_evidence_of_life}
Our potential evidence for DMS in K2-18~b motivates consideration of possible biological activity on the planet. While the present evidence is not as strong as that for CH$_4$ or CO$_2$, upcoming JWST observations of K2-18~b will be able to robustly constrain the presence and abundance of DMS, as discussed in section~\ref{sec:future} and earlier work \citep{Madhusudhan2021}. Here we discuss the plausibility of our DMS abundance constraints from a potential biosphere on K2-18~b in order to inform future observations and retrieval studies.

On Earth, DMS is a by-product of living organisms, with the bulk of the nonanthropogenic DMS in the terrestrial atmosphere being emitted from phytoplankton in marine environments \citep{Charlson1987, Barnes2006}. Both DMS and CH$_3$Cl are thought to be terrestrial biosignatures with no known false positives \citep{catling2018}. On Earth these molecules are produced exclusively by life in relatively small quantities compared to more abundant by-products of life, such as O$_2$, CH$_4$ and N$_2$O; the latter are, therefore, more favoured as biosignatures for Earth-like planets. However, it has been suggested that in H$_2$-rich environments with large biomass, molecules such as DMS and CH$_3$Cl could be abundant and observable for habitable super-Earths \citep{Seager2013} and Hycean worlds \citep{Madhusudhan2021}.

While we infer DMS with marginal confidence, our retrieved DMS abundance spans a relatively wide range across the cases considered, from $\log$(X$_{\text{DMS}}$) = $-4.46 ^{+0.77}_{-0.88}$ at the higher end for the no-offset case to $\log$(X$_{\text{DMS}}$) = $-6.87^{+1.87}_{-3.25}$ for the two-offset case. On Earth, the typical mixing ratios of DMS, found near the ocean surface, are a few hundred parts per trillion \citep{Hopkins2023}. DMS is rapidly depleted at higher altitudes, with a few-day lifetime, due to photochemical reactions with OH and other radicals, ultimately leading to the production of more oxidized sulfur molecules, such as SO$_2$. The upper end of our retrieved abundance for DMS is significantly higher than that on Earth, and we do not detect other sulfur-bearing species; however, the lower end is more plausible \citep{Seager2013}. We note that the infrared absorption cross-sections of DMS are currently limited \citep{dms_cs2_1,dms_cs2_2,HITRAN2016}. New cross-section data may revise our mixing ratio estimates, while future more intensive observations could be sensitive to other sulfur species.

The DMS abundance is also strongly dependent on the chromospheric activity of the host star. A quiescent M-dwarf host star with a lower ultraviolet flux compared to a sun-like star of the same bolometric flux could enable DMS to survive longer and be more abundant in the planetary atmosphere \citep{domagal-goldman2011}. Previous studies have predicted DMS mixing ratios of $\sim$10$^{-7}$--10$^{-6}$ for Earth-like planets and super-Earths orbiting low-activity M dwarfs for plausible biomass estimates \citep{domagal-goldman2011,Seager2013}. However, K2-18~b has been described as a moderately active M3 dwarf \citep{Benneke2019a}, and it may not be quiescent in the extreme-ultraviolet \citep{dosSantos2020}. A high Lyman alpha flux could lead to DMS depletion due to interaction with atomic O produced from CO$_2$ photolysis \citep{Seager2013}. Reaction of DMS with atomic H is significantly slower than with O \citep[see][]{Atkinson2004,Zhang2005}, and the OH abundance would be limited by H$_2$O condensation in K2-18~b.

If the DMS abundance on K2-18~b is indeed confirmed by future observations to be greater than $\sim$10$^{-6}$, that result could require very high biological production rates in the ocean and/or new theoretical developments in our understanding of DMS chemistry (including potential abiotic chemistry) in planets such as K2-18~b. We also note that while our retrievals included a wide array of molecules with strong spectral signatures in the observed wavelength range, future theoretical studies and retrievals could consider an even more expanded set, particularly as accurate cross-section data become available. In particular, other hydrocarbon molecules have similar C-H bands in the 3-3.5 $\mu$m range to that of DMS. Therefore, besides our canonical model, we have explored retrievals with other hydrocarbons such as C$_2$H$_2$, C$_2$H$_6$, CH$_3$OH, HC$_3$N, and hazes with optical properties from \cite{He2023}, which have been predicted to be relevant for Hycean atmospheres \cite[e.g.][]{Tsai2021,Madhusudhan2023}, but we found no definitive evidence for their presence.

Besides DMS, the presence of life could potentially also contribute to the strong chemical disequilibrium indicated by the retrieved atmospheric composition of K2-18~b. For instance, methanogenic bacteria in Earth's oceans are known to be a significant contributor to the atmospheric CH$_4$ budget. It is possible that similar biotic sources may to some extent contribute to the observed CH$_4$ abundance in K2-18~b, if indeed life exists on the planet.

Overall, our findings demonstrate the feasibility of detecting a biosignature molecule in the atmosphere of a habitable-zone sub-Neptune with JWST. This also provides a valuable case study for a framework for biosignature assessment in exoplanets \citep[e.g.][]{catling2018, Meadows2022}. The potential inference of DMS in K2-18~b provides a pathway toward the possible detection of life on an exoplanet with JWST and other current and upcoming large observational facilities. The next steps would involve both (a) more theoretical investigations to understand the possible atmospheric and interior processes at play and (b) more observations to verify the present findings and potentially discover other chemical species.

\subsection{Resolving the Missing Methane Problem}
Our strong detection of CH$_4$ at 5$\sigma$ resolves one of the longest-standing conundrums in exoplanet science --- ``The Missing Methane Problem" \citep{Stevenson2010,madhu_seager2011}. Low-temperature molecules such as CH$_4$ and NH$_3$ are common in the solar system and are seen in the atmospheres of the giant planets \citep{Karkoschka1998, Encrenaz2022, Atreya2016}. These molecules are expected to be  prominent carriers of carbon and nitrogen in H$_2$-rich atmospheres at temperatures below  $\sim$600 K, with H$_2$O being the dominant oxygen carrier \citep{Burrows1999, Lodders2002}. However, no robust detection of CH$_4$ or NH$_3$ has been made in any exoplanetary atmosphere with temperatures below $\sim$800 K, despite atmospheric observations made for several such exoplanets  with HST and Spitzer at wavelengths sensitive to these molecules, e.g. GJ 436~b \citep{Stevenson2010, Knutson2014}, GJ 3470~b \citep{Benneke2019a}, K2-18~b \citep{Benneke2019}, all in the $\sim$300-800 K range \citep[but see][]{Blain2021}.

Atmospheres of temperate sub-Neptunes are expected to exhibit distinct nonequilibrium chemistry, just as in solar system planets. Several processes can cause chemical disequilibrium, including photochemistry, vertical mixing, and volcanic outgassing \citep[e.g.][]{Yung1999}. However, even strong nonequilibrium chemical mechanisms have difficulty explaining the missing CH$_4$ in temperate exoplanetary atmospheres \citep{Line2011, Moses2013}. This missing methane problem has, therefore, remained one of the central puzzles in the area of exoplanetary atmospheres in the pre-JWST era. The present detection of CH$_4$ in K2-18~b, therefore, demonstrates the detectability of CH$_4$ and potentially other hydrocarbons with JWST and opens a new era of atmospheric characterisation of temperate exoplanets in general. The detection suggests that other sub-Neptunes and giant exoplanets with H$_2$-rich atmospheres at similar temperatures may also be conducive for detecting CH$_4$. Such planets therefore represent important targets for homogeneous studies of carbon chemistry in exoplanetary atmospheres, enabling comparative studies with solar system giant planets. 

\subsection{Future Directions}
\label{sec:future}
Our results demonstrate the potential of candidate Hycean worlds as optimal targets in the search for life on exoplanets. These findings motivate further observations and theoretical work to characterise in detail the atmospheric and potential surface conditions of K2-18~b and other candidate Hycean worlds \citep{Madhusudhan2021}. Several upcoming JWST observations of K2-18~b will be able to verify the present findings, in particular more observations with NIRSpec G395H (JWST GO 2372) and MIRI LRS (5-10 $\mu$m) (JWST GO 2722). While the former program can confirm the present findings with higher precision, the latter can specifically confirm the presence of DMS which is expected to have a strong spectral feature around 7 $\mu$m \citep[e.g. Figure 7 of][]{Seager2013}. Such observations are also motivated for a number of other promising candidate Hycean worlds orbiting nearby M dwarfs that are even more favorable to observations than K2-18~b \citep{Madhusudhan2021}. 

Overall, the present results pave the way to a new era of atmospheric characterisation of habitable planets and biosignature detection with JWST. The observations also motivate a wide range of theoretical studies to understand in detail the physical, chemical, and biological conditions on Hycean worlds. Our findings present a first step toward the spectroscopic identification of life beyond the solar system and the assessment of our place in the Universe. 

{\it Acknowledgements:} This work is based on observations made with the NASA/ESA/CSA James Webb Space Telescope as part of Cycle 1 GO Program 2722 (PI: N. Madhusudhan).  We thank NASA, ESA, CSA, STScI, everyone whose efforts have contributed to the JWST, and the exoplanet science community for the thriving current state of the field. This work is supported by research grants to N.M. from the UK Research and Innovation (UKRI) Frontier Grant (EP/X025179/1), the MERAC Foundation, Switzerland, and the UK Science and Technology Facilities Council (STFC) Center for Doctoral Training (CDT) in Data Intensive Science at the University of Cambridge (STFC grant No. ST/P006787/1). N.M. and M.H. acknowledge support from STFC and the MERAC Foundation toward the doctoral studies of M.H. N.M. thanks Tony Roman, Elena Manjavacas, Nestor Espinoza and Sara Kendrew at STScI for their help with planning our JWST observations. J.M. acknowledges support from JWST-GO-02722, which was provided by NASA through a grant from the Space Telescope Science Institute, which is operated by the Association of Universities for Research in Astronomy, Inc., under NASA contract NAS 5-03127. We thank the anonymous referees for their valuable comments on the manuscript. This research has made use of the NASA Exoplanet Archive, which is operated by the California Institute of Technology, under contract with the National Aeronautics and Space Administration under the Exoplanet Exploration Program. This research has made use of the NASA Astrophysics Data System and the Python packages \texttt{NUMPY}, \texttt{SCIPY}, and \texttt{MATPLOTLIB}.

This work was performed using resources provided by the Cambridge Service for Data Driven Discovery operated by the University of Cambridge Research Computing Service (\url{www.csd3.cam.ac.uk}), provided by Dell EMC and Intel using Tier-2 funding from the Engineering and Physical Sciences Research Council (capital grant EP/P020259/1), and DiRAC funding from STFC (\url{www.dirac.ac.uk}).


{\it Author Contributions:} N.M. conceived, planned and led the project. N.M. led the JWST proposal with contributions from S.C., S.S., A.P. and J.M. N.M. and S.S. planned the JWST observations. N.M., M.H. and S.S. conducted the data reduction and analyses. N.M. and S.C. conducted the atmospheric retrievals. N.M., J.M. and A.P. conducted the theoretical interpretation. N.M. led the writing of the manuscript with contributions and comments from all authors.

{\it Data Availability:} Some/all of the data presented in this paper were obtained from the Mikulski Archive for Space Telescopes (MAST) at the Space Telescope Science Institute. The specific observations analyzed can be accessed via \dataset[doi:10.17909/3ds1-8z15]{https://doi.org/10.17909/3ds1-8z15}. The transmission spectra of K2-18 b reported in this work are available on the Open Science Framework, \dataset[doi:10.17605/OSF.IO/36DJH]{https://doi.org/10.17605/OSF.IO/36DJH}.

\facilities{JWST (NIRISS and NIRSpec)}\\

\appendix
\section{Comparison with HST WFC3 Observations}
\label{sec:hst_comparison}
\begin{figure*}
	\includegraphics[width=\textwidth]{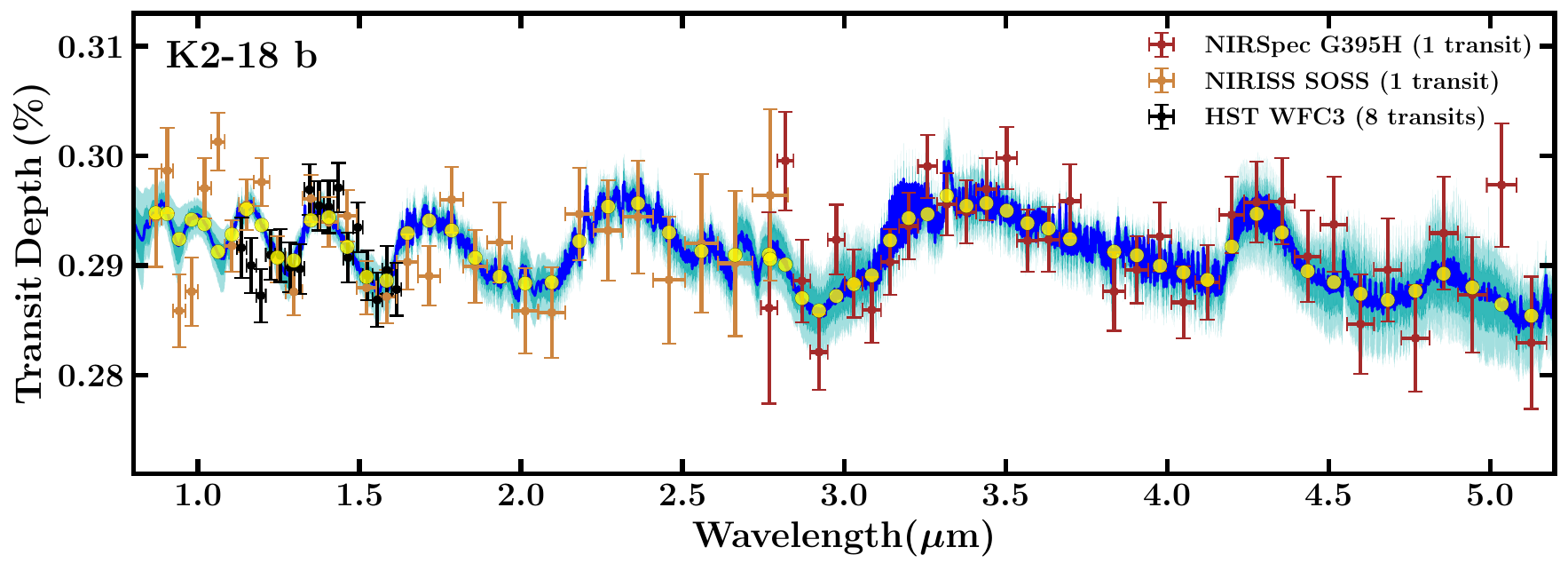}
    \caption{HST and JWST observations of K2-18~b. The black points show prior observations of K2-18~b obtained with HST WFC3 in the 1.1-1.7~$\mu$m range, with data from eight transits, presented by \citet{Benneke2019} using data from HST GO 13665 and GO 14682 programs (PI: B. Benneke). The orange and dark-red points show our JWST NIRISS and NIRSpec observations from one transit each. The data is binned for visual clarity to $R \approx 25$ and $R \approx 55$, respectively, as shown in Fig.~\ref{fig:spectrum}. The dark-blue line denotes the median retrieved spectrum (one-offset case), while medium- and lighter-blue regions denote the 1$\sigma$ and 2$\sigma$ contours, respectively. The yellow points correspond to the median spectrum binned to match the JWST observations. Our JWST NIRISS spectrum is in agreement with the HST WFC3 spectrum for most of the common wavelength range except for two data points toward the blue end of the WFC3 band.}
    \label{fig:wfc3_comparison} 
\end{figure*}

As discussed above, K2-18~b has been previously observed in the 1.1-1.7$\mu$m range with HST WFC3 \citep{Tsiaras2019, Benneke2019}, with these observations leading to inferences of H$_2$O being present in the planet's terminator atmosphere \citep{Tsiaras2019, Benneke2019, Madhusudhan2020} with a high upper limit on CH$_4$ \citep{Madhusudhan2020}. The spectrum has also been suggested to be explained by CH$_4$ or a combination of CH$_4$ and H$_2$O, instead of H$_2$O alone, due to the degeneracy between the two molecules in the WFC3 band \citep{Blain2021}. Fig. \ref{fig:wfc3_comparison} overlays these prior observations presented by \citet{Benneke2019} with our new JWST spectrum. It can be seen that the new NIRISS data presented in this work is in good agreement with the prior HST WFC3 data in general, with the notable exception of two data points in the blue end of the WFC3 band, which show a 2-3$\sigma$ deviation from the NIRISS observations and corresponding spectral fit. The two deviant WFC3 points are inconsistent with a CH$_4$ absorption peak in the retrieved spectral fit. As such, it is possible that prior inferences of H$_2$O over CH$_4$ were affected by these two points. 

\section{Comparison with Retrievals Assuming No Clouds}
\label{sec:no_clouds}
\begin{figure*}
	\includegraphics[width=\textwidth]{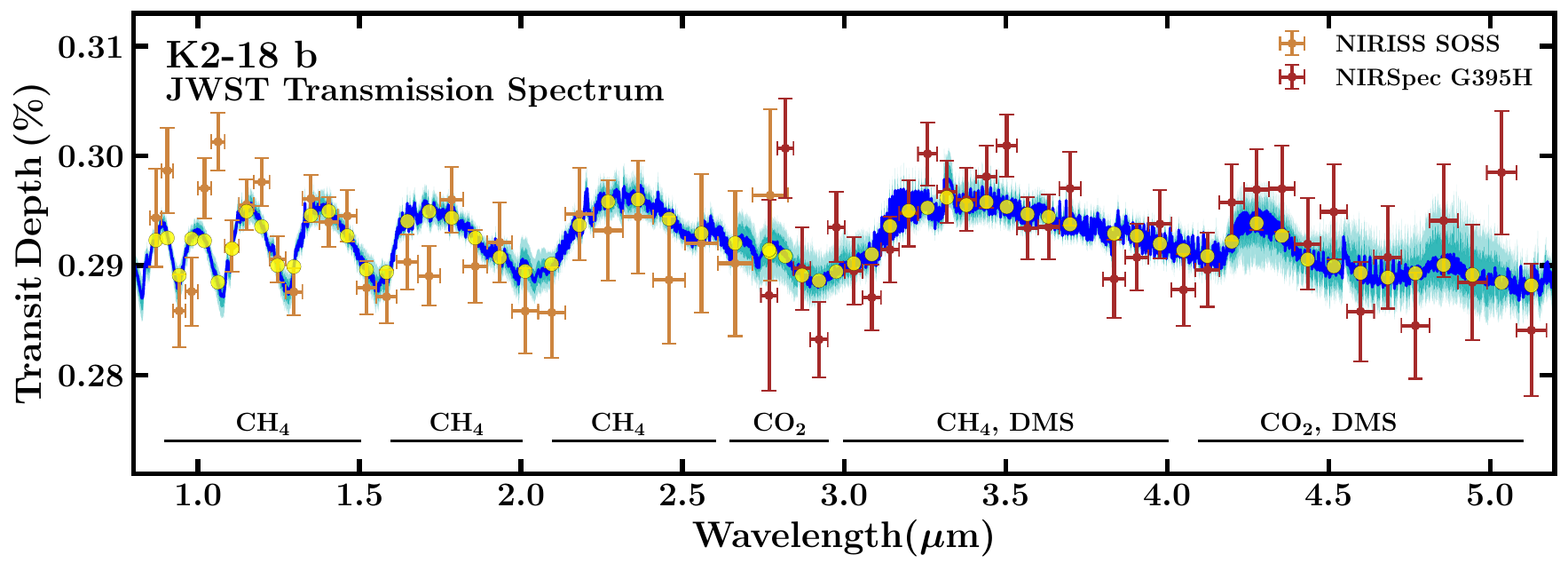}
    \caption{
Retrieved spectral fit of the JWST transmission spectrum of K2-18b using a model without clouds/hazes in the one-offset case. Details of the figure are the same as for Fig. \ref{fig:spectrum} except that the retrieval does not include clouds/hazes in the model. The retrieval produces a comparable fit to the cloudy case (Fig. \ref{fig:spectrum}) across most of the wavelength range, differing mostly towards the bluer end of the NIRISS band, as discussed in Appendix~\ref{sec:no_clouds}. The NIRSpec spectrum is vertically offset by $-30$ ppm, corresponding to the median retrieved offset.
     }
    \label{fig:spec_no_clouds} 
\end{figure*}

As discussed in section~\ref{sec:clouds} our retrievals with the canonical model provide nominal constraints on the properties of possible clouds/hazes. Here we show a retrieved spectral fit to our JWST transmission spectrum using a model with no clouds/hazes (Fig. \ref{fig:spec_no_clouds}). This is applied to the one-offset case discussed in section~\ref{sec:retrieval}, with all other factors being the same as for the canonical retrieval with clouds/hazes included. While the Bayesian evidence is higher for the canonical model, as discussed in section~\ref{sec:clouds}, the no clouds/hazes model provides a reasonable fit (Fig. \ref{fig:spec_no_clouds}) to most of the observed spectrum. It differs from the canonical model fit (Fig. \ref{fig:spectrum}) only toward the blue end of the NIRISS spectrum. The retrieved abundances in the present case are also consistent with the canonical case within the 1$\sigma$ uncertainties. Therefore, further observations in the optical are needed to more robustly constrain the presence and properties of clouds/hazes in the atmosphere of K2-18~b. 

\section{Bayesian Priors for Atmospheric Retrieval}
\label{sec:priors}
Table \ref{tab:retrieval_priors} shows the Bayesian prior probability distributions used in the retrievals presented in this work. All but the last six parameters correspond to the canonical model described in section \ref{sec:retrieval_setup}. $\delta_\mathrm{Nirspec}$ denotes the linear offset applied to the NIRSpec G395H data in the 1-offset case in parts-per-million, while $\delta_\mathrm{NRS1}$ and $\delta_\mathrm{NRS2}$ denote the separate offsets applied to observations from the NIRSpec NRS1 and NRS2 detectors, respectively, in parts-per-million. $T_\mathrm{phot}$, $T_\mathrm{het}$ and $f_\mathrm{het}$ are the photospheric temperature, starspot/faculae (i.e. heterogeneity) temperature, and coverage fraction, respectively, and are used in the unocculted stellar heterogeneities modelling discussed in section \ref{sec:stellar_heterogeneities}. $\mathcal{U}(X, Y)$ denotes a uniform probability distribution between X and Y, while $\mathcal{N}(\mu, \sigma^2)$ denotes a normal distribution of mean $\mu$ and variance $\sigma^2$. For any additional chemical species considered, such as the hydrocarbons described in section \ref{sec:possible_evidence_of_life}, we use the same uniform mixing ratio priors as those shown in table \ref{tab:retrieval_priors}.

\begin{table*}
\centering
\begin{tabular}{l|c|l}
    Parameter & Bayesian Prior  & Description\\[0.5mm]
    \hline
    \hline
    $\mathrm{log}(X_\mathrm{H_2O})$& $\mathcal{U}$(-12, -0.3) & Mixing ratio of H$_2$O \\[0.5mm]
    $\mathrm{log}(X_\mathrm{CH_4})$  & $\mathcal{U}$(-12, -0.3) & Mixing ratio of CH$_4$\\[0.5mm]
    $\mathrm{log}(X_\mathrm{NH_3})$  & $\mathcal{U}$(-12, -0.3) & Mixing ratio of NH$_3$ \\[0.5mm]
    $\mathrm{log}(X_\mathrm{HCN})$  & $\mathcal{U}$(-12, -0.3)& Mixing ratio of HCN\\[0.5mm]
    $\mathrm{log}(X_\mathrm{CO})$ & $\mathcal{U}$(-12, -0.3) & Mixing ratio of CO\\[0.5mm]
    $\mathrm{log}(X_\mathrm{CO_2})$& $\mathcal{U}$(-12, -0.3) & Mixing ratio of CO$_2$\\[0.5mm]
    $\mathrm{log}(X_\mathrm{DMS})$  & $\mathcal{U}$(-12, -0.3) & Mixing ratio of DMS\\[0.5mm]
    $\mathrm{log}(X_\mathrm{CS_2})$& $\mathcal{U}$(-12, -0.3) & Mixing ratio CS$_2$\\[0.5mm]
    $\mathrm{log}(X_\mathrm{CH_3Cl})$  & $\mathcal{U}$(-12, -0.3)& Mixing ratio of CH$_3$Cl\\[0.5mm]
    $\mathrm{log}(X_\mathrm{OCS})$ & $\mathcal{U}$(-12, -0.3)& Mixing ratio of OCS\\[0.5mm]
    $\mathrm{log}(X_\mathrm{N_2O})$ & $\mathcal{U}$(-12, -0.3) & Mixing ratio N$_2$O \\[0.5mm]
    $T_0 / \mathrm{K} $ & $\mathcal{U}$(0, 500)& Temperature at 1$\mu$bar  \\[0.5mm]
    $\alpha_1 / \mathrm{K}^{-\frac{1}{2}}$  & $\mathcal{U}$(0.02, 2.00) & P-T profile curvature \\[0.5mm]
    $\alpha_2/ \mathrm{K}^{-\frac{1}{2}}$& $\mathcal{U}$(0.02, 2.00)  & P-T profile curvature \\[0.5mm]

    $\mathrm{log}(P_1/\mathrm{bar})$   & $\mathcal{U}$(-6, 0) & P-T profile region limit\\[0.5mm]
    $\mathrm{log}(P_2/\mathrm{bar})$  & $\mathcal{U}$(-6, 0) & P-T profile region limit \\[0.5mm]
    $\mathrm{log}(P_3/\mathrm{bar})$   & $\mathcal{U}$(-2, 0)& P-T profile region limit \\[0.5mm]
    $\mathrm{log}(P_\mathrm{ref}/\mathrm{bar})$   & $\mathcal{U}$(-6, 0) & Reference pressure at R$_\mathrm{P}$\\[0.5mm]
    $\mathrm{log}(a)$  & $\mathcal{U}$(-4, 10)& Rayleigh enhancement factor \\[0.5mm]
    $\gamma$   & $\mathcal{U}$(-20, 2)& Scattering slope \\[0.5mm]
    $\mathrm{log}(P_\mathrm{c}/\mathrm{bar})$  & $\mathcal{U}$(-6, 1)& Cloud top pressure \\[0.5mm]
    $\phi$  & $\mathcal{U}$(0, 1)& Cloud/haze coverage fraction\\[0.5mm]
    $\delta_\mathrm{Nirspec} / \mathrm{ppm}$  & $\mathcal{U}$(-100, 100)& NIRSpec dataset offset \\[0.5mm]
    $\delta_\mathrm{NRS1} / \mathrm{ppm}$  & $\mathcal{U}$(-100, 100)& NIRSpec NRS1 dataset offset \\[0.5mm]
    $\delta_\mathrm{NRS2} / \mathrm{ppm}$ & $\mathcal{U}$(-100, 100) & NIRSpec NRS2 dataset offset \\[0.5mm]
    $T_\mathrm{phot} /$  K  & $\mathcal{N}(3457, 100^2)$& Stellar photosphere temperature\\[0.5mm]
    $T_\mathrm{het} /$  K & $\mathcal{U}(1729, 4148)$ & Stellar heterogeneity temperature \\[0.5mm]
    $f_\mathrm{het} $  & $\mathcal{U}$(0, 0.5)& Stellar Heterogeneity coverage fraction\\[0.5mm]
     \hline
     
\end{tabular}
\caption{Model parameters and corresponding Bayesian prior probability distributions used in the atmospheric retrievals performed in this work. All but the last six parameters are used in the canonical atmospheric model described in section \ref{sec:retrieval_setup}.}
\label{tab:retrieval_priors}
\end{table*}

\bibliography{ms.bib,references.bib}

\begin{thebibliography}{}
\expandafter\ifx\csname natexlab\endcsname\relax\def\natexlab#1{#1}\fi
\providecommand{\url}[1]{\href{#1}{#1}}
\providecommand{\dodoi}[1]{doi:~\href{http://doi.org/#1}{\nolinkurl{#1}}}
\providecommand{\doeprint}[1]{\href{http://ascl.net/#1}{\nolinkurl{http://ascl.net/#1}}}
\providecommand{\doarXiv}[1]{\href{https://arxiv.org/abs/#1}{\nolinkurl{https://arxiv.org/abs/#1}}}

\bibitem[{{Abel} {et~al.}(2011){Abel}, {Frommhold}, {Li}, \& {Hunt}}]{abel2011}
{Abel}, M., {Frommhold}, L., {Li}, X., \& {Hunt}, K. L.~C. 2011, J. Phys. Chem. A, 115, 6805, \dodoi{10.1021/jp109441f}

\bibitem[{Albert {et~al.}(2023)Albert, Lafrenière, Doyon, Étienne Artigau, Volk, Goudfrooij, Martel, Radica, Rowe, Espinoza, Roy, Filippazzo, Darveau-Bernier, Talens, Sivaramakrishnan, Willott, Fullerton, LaMassa, Hutchings, Rowlands, Vila, Zhou, Aldridge, Maszkiewicz, Beaulieu, Cook, Piaulet, Roy, Lamontagne, Morel, Frost, Salhi, Coulombe, Benneke, MacDonald, Johnstone, Turner, Fournier-Tondreau, Allart, \& Kaltenegger}]{Albert2023}
Albert, L., Lafrenière, D., Doyon, R., {et~al.} 2023, \pasp, 135, 075001, \dodoi{10.1088/1538-3873/acd7a3}

\bibitem[{{Alderson} {et~al.}(2023){Alderson}, {Wakeford}, {Alam}, {Batalha}, {Lothringer}, {Adams Redai}, {Barat}, {Brande}, {Damiano}, {Daylan}, {Espinoza}, {Flagg}, {Goyal}, {Grant}, {Hu}, {Inglis}, {Lee}, {Mikal-Evans}, {Ramos-Rosado}, {Roy}, {Wallack}, {Batalha}, {Bean}, {Benneke}, {Berta-Thompson}, {Carter}, {Changeat}, {Col{\'o}n}, {Crossfield}, {D{\'e}sert}, {Foreman-Mackey}, {Gibson}, {Kreidberg}, {Line}, {L{\'o}pez-Morales}, {Molaverdikhani}, {Moran}, {Morello}, {Moses}, {Mukherjee}, {Schlawin}, {Sing}, {Stevenson}, {Taylor}, {Aggarwal}, {Ahrer}, {Allen}, {Barstow}, {Bell}, {Blecic}, {Casewell}, {Chubb}, {Crouzet}, {Cubillos}, {Decin}, {Feinstein}, {Fortney}, {Harrington}, {Heng}, {Iro}, {Kempton}, {Kirk}, {Knutson}, {Krick}, {Leconte}, {Lendl}, {MacDonald}, {Mancini}, {Mansfield}, {May}, {Mayne}, {Miguel}, {Nikolov}, {Ohno}, {Palle}, {Parmentier}, {Petit dit de la Roche}, {Piaulet}, {Powell}, {Rackham}, {Redfield}, {Rogers}, {Rustamkulov}, {Tan}, {Tremblin}, {Tsai}, {Turner}, {de Val-Borro},
  {Venot}, {Welbanks}, {Wheatley}, \& {Zhang}}]{Alderson2023}
{Alderson}, L., {Wakeford}, H.~R., {Alam}, M.~K., {et~al.} 2023, \nat, 614, 664, \dodoi{10.1038/s41586-022-05591-3}

\bibitem[{{Atkinson} {et~al.}(2004){Atkinson}, {Baulch}, {Cox}, {Crowley}, {Hampson}, {Hynes}, {Jenkin}, {Rossi}, \& {Troe}}]{Atkinson2004}
{Atkinson}, R., {Baulch}, D.~L., {Cox}, R.~A., {et~al.} 2004, Atmospheric Chemistry \& Physics, 4, 1461, \dodoi{10.5194/acp-4-1461-2004}

\bibitem[{Atreya {et~al.}(2018)Atreya, Crida, Guillot, Lunine, Madhusudhan, \& Mousis}]{Atreya2016}
Atreya, S.~K., Crida, A., Guillot, T., {et~al.} 2018, The Origin and Evolution of Saturn, with Exoplanet Perspective, Cambridge Planetary Science (Cambridge University Press), 5–43, \dodoi{10.1017/9781316227220.002}

\bibitem[{{August} {et~al.}(2023){August}, {Bean}, {Zhang}, {Lunine}, {Xue}, {Line}, \& {Smith}}]{August2023}
{August}, P.~C., {Bean}, J.~L., {Zhang}, M., {et~al.} 2023, arXiv e-prints, arXiv:2305.07753, \dodoi{10.48550/arXiv.2305.07753}

\bibitem[{Auwera \& Fayt(2006)}]{ocs_4}
Auwera, J.~V., \& Fayt, A. 2006, Journal of Molecular Structure, 780-781, 134, \dodoi{10.1016/j.molstruc.2005.04.052}

\bibitem[{{Barber} {et~al.}(2014){Barber}, {Strange}, {Hill}, {Polyansky}, {Mellau}, {Yurchenko}, \& {Tennyson}}]{barber2014}
{Barber}, R.~J., {Strange}, J.~K., {Hill}, C., {et~al.} 2014, \mnras, 437, 1828, \dodoi{10.1093/mnras/stt2011}

\bibitem[{{Barclay} {et~al.}(2021)}]{Barclay2021}
{Barclay}, T., {et~al.} 2021, Astron. J., 162, 300, \dodoi{10.3847/1538-3881/ac2824}

\bibitem[{{Barnes} {et~al.}(2006){Barnes}, {Hjorth}, \& {Mihalopoulos}}]{Barnes2006}
{Barnes}, I., {Hjorth}, J., \& {Mihalopoulos}, N. 2006, Chem. Rev., 106, 940, \dodoi{10.1021/cr020529+}

\bibitem[{{B{\'e}ky} {et~al.}(2014){B{\'e}ky}, {Kipping}, \& {Holman}}]{spotrod2014}
{B{\'e}ky}, B., {Kipping}, D.~M., \& {Holman}, M.~J. 2014, \mnras, 442, 3686, \dodoi{10.1093/mnras/stu1061}

\bibitem[{{Benneke} \& {Seager}(2013)}]{Benneke2013}
{Benneke}, B., \& {Seager}, S. 2013, \apj, 778, 153, \dodoi{10.1088/0004-637X/778/2/153}

\bibitem[{{Benneke} {et~al.}(2019{\natexlab{a}}){Benneke}, {Wong}, {Piaulet}, {Knutson}, {Lothringer}, {Morley}, {Crossfield}, {Gao}, {Greene}, {Dressing}, {Dragomir}, {Howard}, {McCullough}, {Kempton}, {Fortney}, \& {Fraine}}]{Benneke2019}
{Benneke}, B., {Wong}, I., {Piaulet}, C., {et~al.} 2019{\natexlab{a}}, \apjl, 887, L14, \dodoi{10.3847/2041-8213/ab59dc}

\bibitem[{{Benneke} {et~al.}(2019{\natexlab{b}}){Benneke}, {Knutson}, {Lothringer}, {Crossfield}, {Moses}, {Morley}, {Kreidberg}, {Fulton}, {Dragomir}, {Howard}, {Wong}, {D{\'e}sert}, {McCullough}, {Kempton}, {Fortney}, {Gilliland}, {Deming}, \& {Kammer}}]{Benneke2019a}
{Benneke}, B., {Knutson}, H.~A., {Lothringer}, J., {et~al.} 2019{\natexlab{b}}, Nature Astronomy, 3, 813, \dodoi{10.1038/s41550-019-0800-5}

\bibitem[{{B{\'e}zard} {et~al.}(2022){B{\'e}zard}, {Charnay}, \& {Blain}}]{Bezard2022}
{B{\'e}zard}, B., {Charnay}, B., \& {Blain}, D. 2022, Nat. Astron., 6, 537, \dodoi{10.1038/s41550-022-01678-z}

\bibitem[{{Birkmann} {et~al.}(2014)}]{Birkmann2014}
{Birkmann}, S.~M., {et~al.} 2014, in Society of Photo-Optical Instrumentation Engineers (SPIE) Conference Series, Vol. 9143, Space Telescopes and Instrumentation 2014: Optical, Infrared, and Millimeter Wave, ed. J.~{Oschmann}, Jacobus~M., M.~{Clampin}, G.~G. {Fazio}, \& H.~A. {MacEwen}, 914308, \dodoi{10.1117/12.2054642}

\bibitem[{{Blain} {et~al.}(2021){Blain}, {Charnay}, \& {B{\'e}zard}}]{Blain2021}
{Blain}, D., {Charnay}, B., \& {B{\'e}zard}, B. 2021, Astron. Astrophys., 646, A15, \dodoi{10.1051/0004-6361/202039072}

\bibitem[{{Borysow} {et~al.}(1988){Borysow}, {Frommhold}, \& {Birnbaum}}]{borysow1988}
{Borysow}, J., {Frommhold}, L., \& {Birnbaum}, G. 1988, \apj, 326, 509, \dodoi{10.1086/166112}

\bibitem[{Bouanich {et~al.}(1986)Bouanich, Blanquet, Walrand, \& Courtoy}]{ocs_1}
Bouanich, J.-P., Blanquet, G., Walrand, J., \& Courtoy, C.~P. 1986, Journal of Quantitative Spectroscopy and Radiative Transfer, 36, 295, \dodoi{10.1016/0022-4073(86)90053-1}

\bibitem[{Bray {et~al.}(2011)Bray, Perrin, Jacquemart, \& Lacome}]{ch3cl_1}
Bray, C., Perrin, A., Jacquemart, D., \& Lacome, N. 2011, Journal of Quantitative Spectroscopy and Radiative Transfer, 112, 2446, \dodoi{10.1016/j.jqsrt.2011.06.018}

\bibitem[{{Buchner} {et~al.}(2014){Buchner}, {Georgakakis}, {Nandra}, {Hsu}, {Rangel}, {Brightman}, {Merloni}, {Salvato}, {Donley}, \& {Kocevski}}]{Buchner2014}
{Buchner}, J., {Georgakakis}, A., {Nandra}, K., {et~al.} 2014, \aap, 564, A125, \dodoi{10.1051/0004-6361/201322971}

\bibitem[{{Burrows} \& {Sharp}(1999)}]{Burrows1999}
{Burrows}, A., \& {Sharp}, C.~M. 1999, \apj, 512, 843, \dodoi{10.1086/306811}

\bibitem[{{Bushouse}(2020)}]{Bushouse2020}
{Bushouse}, H. 2020, in Astronomical Society of the Pacific Conference Series, Vol. 527, Astronomical Data Analysis Software and Systems XXIX, ed. R.~{Pizzo}, E.~R. {Deul}, J.~D. {Mol}, J.~{de Plaa}, \& H.~{Verkouter}, 583

\bibitem[{{Catling} {et~al.}(2018){Catling}, {Krissansen-Totton}, {Kiang}, {Crisp}, {Robinson}, {DasSarma}, {Rushby}, {Del Genio}, {Bains}, \& {Domagal-Goldman}}]{catling2018}
{Catling}, D.~C., {Krissansen-Totton}, J., {Kiang}, N.~Y., {et~al.} 2018, Astrobiology, 18, 709, \dodoi{10.1089/ast.2017.1737}

\bibitem[{{Charlson} {et~al.}(1987){Charlson}, {Warren}, {Lovelock}, \& {Andreae}}]{Charlson1987}
{Charlson}, R.~J., {Warren}, S.~G., {Lovelock}, J.~E., \& {Andreae}, M.~O. 1987, \nat, 326, 655, \dodoi{10.1038/326655a0}

\bibitem[{{Cloutier} {et~al.}(2017){Cloutier}, {Astudillo-Defru}, {Doyon}, {Bonfils}, {Almenara}, {Benneke}, {Bouchy}, {Delfosse}, {Ehrenreich}, {Forveille}, {Lovis}, {Mayor}, {Menou}, {Murgas}, {Pepe}, {Rowe}, {Santos}, {Udry}, \& {W{\"u}nsche}}]{cloutier2017}
{Cloutier}, R., {Astudillo-Defru}, N., {Doyon}, R., {et~al.} 2017, \aap, 608, A35, \dodoi{10.1051/0004-6361/201731558}

\bibitem[{{Cloutier} {et~al.}(2019)}]{Cloutier2019}
{Cloutier}, R., {et~al.} 2019, Astron. Astrophys., 621, A49, \dodoi{10.1051/0004-6361/201833995}

\bibitem[{{Coles} {et~al.}(2019){Coles}, {Yurchenko}, \& {Tennyson}}]{Coles2019}
{Coles}, P.~A., {Yurchenko}, S.~N., \& {Tennyson}, J. 2019, \mnras, 490, 4638, \dodoi{10.1093/mnras/stz2778}

\bibitem[{{Constantinou} \& {Madhusudhan}(2022)}]{Constantinou2022}
{Constantinou}, S., \& {Madhusudhan}, N. 2022, Mon. Notices Royal Astron. Soc., 514, 2073, \dodoi{10.1093/mnras/stac1277}

\bibitem[{{Constantinou} {et~al.}(2023){Constantinou}, {Madhusudhan}, \& {Gandhi}}]{Constantinou2023}
{Constantinou}, S., {Madhusudhan}, N., \& {Gandhi}, S. 2023, \apjl, 943, L10, \dodoi{10.3847/2041-8213/acaead}

\bibitem[{Coulombe {et~al.}(2023)Coulombe, Benneke, Challener, Piette, Wiser, Mansfield, MacDonald, Beltz, Feinstein, Radica, Savel, Santos, Bean, Parmentier, Wong, Rauscher, Komacek, Kempton, Tan, Hammond, Lewis, Line, Lee, Shivkumar, Crossfield, Nixon, Rackham, Wakeford, Welbanks, Zhang, Batalha, Berta-Thompson, Changeat, Désert, Espinoza, Goyal, Harrington, Knutson, Kreidberg, López-Morales, Shporer, Sing, Stevenson, Aggarwal, Ahrer, Alam, Bell, Blecic, Caceres, Carter, Casewell, Crouzet, Cubillos, Decin, Fortney, Gibson, Heng, Henning, Iro, Kendrew, Lagage, Leconte, Lendl, Lothringer, Mancini, Mikal-Evans, Molaverdikhani, Nikolov, Ohno, Palle, Piaulet, Redfield, Roy, Tsai, Venot, \& Wheatley}]{coulombe_broadband_2023}
Coulombe, L.-P., Benneke, B., Challener, R., {et~al.} 2023, A broadband thermal emission spectrum of the ultra-hot {Jupiter} {WASP}-18b,  arXiv.
\newblock \url{http://arxiv.org/abs/2301.08192}

\bibitem[{{Csizmadia} {et~al.}(2013){Csizmadia}, {Pasternacki}, {Dreyer}, {Cabrera}, {Erikson}, \& {Rauer}}]{Csizmadia2013}
{Csizmadia}, S., {Pasternacki}, T., {Dreyer}, C., {et~al.} 2013, \aap, 549, A9, \dodoi{10.1051/0004-6361/201219888}

\bibitem[{{Czesla} {et~al.}(2009){Czesla}, {Huber}, {Wolter}, {Schr{\"o}ter}, \& {Schmitt}}]{Czesla2009}
{Czesla}, S., {Huber}, K.~F., {Wolter}, U., {Schr{\"o}ter}, S., \& {Schmitt}, J.~H.~M.~M. 2009, \aap, 505, 1277, \dodoi{10.1051/0004-6361/200912454}

\bibitem[{Daumont {et~al.}(2001)Daumont, Auwera, Teffo, Perevalov, \& Tashkun}]{n2o_2}
Daumont, L., Auwera, J., Teffo, J.-L., Perevalov, V., \& Tashkun, S. 2001, Journal of Molecular Spectroscopy, 208, 281, \dodoi{10.1006/jmsp.2001.8400}

\bibitem[{{Domagal-Goldman} {et~al.}(2011){Domagal-Goldman}, {Meadows}, {Claire}, \& {Kasting}}]{domagal-goldman2011}
{Domagal-Goldman}, S.~D., {Meadows}, V.~S., {Claire}, M.~W., \& {Kasting}, J.~F. 2011, Astrobiology, 11, 419, \dodoi{10.1089/ast.2010.0509}

\bibitem[{{dos Santos} {et~al.}(2020){dos Santos}, {Ehrenreich}, {Bourrier}, {Astudillo-Defru}, {Bonfils}, {Forget}, {Lovis}, {Pepe}, \& {Udry}}]{dosSantos2020}
{dos Santos}, L.~A., {Ehrenreich}, D., {Bourrier}, V., {et~al.} 2020, \aap, 634, L4, \dodoi{10.1051/0004-6361/201937327}

\bibitem[{Doyon {et~al.}(2012)Doyon, Hutchings, Beaulieu, Albert, Lafreni{\`e}re, Willott, Touahri, Rowlands, Maszkiewicz, Fullerton, Volk, Martel, Chayer, Sivaramakrishnan, Abraham, Ferrarese, Jayawardhana, Johnstone, Meyer, Pipher, \& Sawicki}]{doyon_jwst_2012}
Doyon, R., Hutchings, J.~B., Beaulieu, M., {et~al.} 2012, in Space Telescopes and Instrumentation 2012: Optical, Infrared, and Millimeter Wave, ed. M.~C. Clampin, G.~G. Fazio, H.~A. MacEwen, \& J.~M.~O. Jr., Vol. 8442, International Society for Optics and Photonics (SPIE), 84422R, \dodoi{10.1117/12.926578}

\bibitem[{{Doyon} {et~al.}(2023){Doyon}, {Willott}, {Hutchings}, {Sivaramakrishnan}, {Albert}, {Lafreniere}, {Rowlands}, {Begona Vila}, {Martel}, {LaMassa}, {Aldridge}, {Artigau}, {Cameron}, {Chayer}, {Cook}, {Cooper}, {Darveau-Bernier}, {Dupuis}, {Earnshaw}, {Espinoza}, {Filippazzo}, {Fullerton}, {Gaudreau}, {Gawlik}, {Goudfrooij}, {Haley}, {Kammerer}, {Kendall}, {Lambros}, {Ilinca Ignat}, {Maszkiewicz}, {McColgan}, {Morishita}, {Ouellette}, {Pacifici}, {Philippi}, {Radica}, {Ravindranath}, {Rowe}, {Roy}, {Saad}, {Sohn}, {Talens}, {Thatte}, {Taylor}, {Vandal}, {Volk}, {Wander}, {Warner}, {Zheng}, {Zhou}, {Abraham}, {Beaulieu}, {Benneke}, {Ferrarese}, {Johnstone}, {Kaltenegger}, {Meyer}, {Pipher}, {Rameau}, {Rieke}, {Salhi}, \& {Sawicki}}]{Doyon2023}
{Doyon}, R., {Willott}, C.~J., {Hutchings}, J.~B., {et~al.} 2023, arXiv e-prints, arXiv:2306.03277, \dodoi{10.48550/arXiv.2306.03277}

\bibitem[{{Encrenaz}(2022)}]{Encrenaz2022}
{Encrenaz}, T. 2022, Research in Astronomy and Astrophysics, 22, 122001, \dodoi{10.1088/1674-4527/ac97d1}

\bibitem[{Espinoza \& Jordán(2015)}]{espinoza_limb_2015}
Espinoza, N., \& Jordán, A. 2015, Monthly Notices of the Royal Astronomical Society, 450, 1879, \dodoi{10.1093/mnras/stv744}

\bibitem[{{Espinoza} {et~al.}(2019){Espinoza}, {Rackham}, {Jord{\'a}n}, {Apai}, {L{\'o}pez-Morales}, {Osip}, {Grimm}, {Hoeijmakers}, {Wilson}, {Bixel}, {McGruder}, {Rodler}, {Weaver}, {Lewis}, {Fortney}, \& {Fraine}}]{Espinoza2019}
{Espinoza}, N., {Rackham}, B.~V., {Jord{\'a}n}, A., {et~al.} 2019, \mnras, 482, 2065, \dodoi{10.1093/mnras/sty2691}

\bibitem[{{Feinstein} {et~al.}(2023){Feinstein}, {Radica}, {Welbanks}, {Murray}, {Ohno}, {Coulombe}, {Espinoza}, {Bean}, {Teske}, {Benneke}, {Line}, {Rustamkulov}, {Saba}, {Tsiaras}, {Barstow}, {Fortney}, {Gao}, {Knutson}, {MacDonald}, {Mikal-Evans}, {Rackham}, {Taylor}, {Parmentier}, {Batalha}, {Berta-Thompson}, {Carter}, {Changeat}, {dos Santos}, {Gibson}, {Goyal}, {Kreidberg}, {L{\'o}pez-Morales}, {Lothringer}, {Miguel}, {Molaverdikhani}, {Moran}, {Morello}, {Mukherjee}, {Sing}, {Stevenson}, {Wakeford}, {Ahrer}, {Alam}, {Alderson}, {Allen}, {Batalha}, {Bell}, {Blecic}, {Brande}, {Caceres}, {Casewell}, {Chubb}, {Crossfield}, {Crouzet}, {Cubillos}, {Decin}, {D{\'e}sert}, {Harrington}, {Heng}, {Henning}, {Iro}, {Kempton}, {Kendrew}, {Kirk}, {Krick}, {Lagage}, {Lendl}, {Mancini}, {Mansfield}, {May}, {Mayne}, {Nikolov}, {Palle}, {Petit dit de la Roche}, {Piaulet}, {Powell}, {Redfield}, {Rogers}, {Roman}, {Roy}, {Nixon}, {Schlawin}, {Tan}, {Tremblin}, {Turner}, {Venot}, {Waalkes}, {Wheatley}, \&
  {Zhang}}]{Feinstein2023}
{Feinstein}, A.~D., {Radica}, M., {Welbanks}, L., {et~al.} 2023, \nat, 614, 670, \dodoi{10.1038/s41586-022-05674-1}

\bibitem[{Feroz {et~al.}(2009)Feroz, Hobson, \& Bridges}]{Feroz2009}
Feroz, F., Hobson, M.~P., \& Bridges, M. 2009, Monthly Notices of the Royal Astronomical Society, 398, 1601, \dodoi{10.1111/j.1365-2966.2009.14548.x}

\bibitem[{{Ferruit} {et~al.}(2012)}]{ferruit2012}
{Ferruit}, P., {et~al.} 2012, in Society of Photo-Optical Instrumentation Engineers (SPIE) Conference Series, Vol. 8442, Space Telescopes and Instrumentation 2012: Optical, Infrared, and Millimeter Wave, ed. M.~C. {Clampin}, G.~G. {Fazio}, H.~A. {MacEwen}, \& J.~{Oschmann}, Jacobus~M., 84422O, \dodoi{10.1117/12.925810}

\bibitem[{{Fulton} \& {Petigura}(2018)}]{Fulton2018}
{Fulton}, B.~J., \& {Petigura}, E.~A. 2018, AJ, 156, 264, \dodoi{10.3847/1538-3881/aae828}

\bibitem[{{Gandhi} \& {Madhusudhan}(2017)}]{gandhi2017}
{Gandhi}, S., \& {Madhusudhan}, N. 2017, Mon. Notices Royal Astron. Soc., 472, 2334, \dodoi{10.1093/mnras/stx1601}

\bibitem[{{Gandhi} {et~al.}(2020){Gandhi}, {Brogi}, {Yurchenko}, {Tennyson}, {Coles}, {Webb}, {Birkby}, {Guilluy}, {Hawker}, {Madhusudhan}, {Bonomo}, \& {Sozzetti}}]{gandhi2020}
{Gandhi}, S., {Brogi}, M., {Yurchenko}, S.~N., {et~al.} 2020, \mnras, 495, 224, \dodoi{10.1093/mnras/staa981}

\bibitem[{{Golebiowski} {et~al.}(2014){Golebiowski}, {de Ghellinck d'Elseghem Vaernewijck}, {Herman}, {Vander Auwera}, \& {Fayt}}]{ocs_2}
{Golebiowski}, D., {de Ghellinck d'Elseghem Vaernewijck}, X., {Herman}, M., {Vander Auwera}, J., \& {Fayt}, A. 2014, \jqsrt, 149, 184, \dodoi{10.1016/j.jqsrt.2014.07.005}

\bibitem[{Gordon {et~al.}(2017)Gordon, Rothman, Hill, Kochanov, Tan, Bernath, Birk, Boudon, Campargue, Chance, Drouin, Flaud, Gamache, Hodges, Jacquemart, Perevalov, Perrin, Shine, Smith, Tennyson, Toon, H.~Tran, Barbe, Csaszar, Devi, Furtenbacher, Harrison, Hartmann, Jolly, Johnson, Karman, Kleiner, Kyuberis, Loos, Lyulin, Massie, Mikhailenko, Moazzen-Ahmadi, Muller, Naumenko, Nikitin, Polyansky, Rey, Rotger, Sharpe, Sung, Starikova, Tashkun, Auwera, Wagner, Wilzewski, Wcislo, Yu, \& Zak}]{HITRAN2016}
Gordon, I., Rothman, L., Hill, C., {et~al.} 2017, Journal of Quantitative Spectroscopy and Radiative Transfer, \dodoi{10.1016/j.jqsrt.2017.06.038}

\bibitem[{{Hargreaves} {et~al.}(2020){Hargreaves}, {Gordon}, {Rey}, {Nikitin}, {Tyuterev}, {Kochanov}, \& {Rothman}}]{Hargreaves2020}
{Hargreaves}, R.~J., {Gordon}, I.~E., {Rey}, M., {et~al.} 2020, \apjs, 247, 55, \dodoi{10.3847/1538-4365/ab7a1a}

\bibitem[{{Harris} {et~al.}(2006){Harris}, {Tennyson}, {Kaminsky}, {Pavlenko}, \& {Jones}}]{harris2006}
{Harris}, G.~J., {Tennyson}, J., {Kaminsky}, B.~M., {Pavlenko}, Y.~V., \& {Jones}, H.~R.~A. 2006, \mnras, 367, 400, \dodoi{10.1111/j.1365-2966.2005.09960.x}

\bibitem[{{He} {et~al.}(2023){He}, {Radke}, {Moran}, {Horst}, {Lewis}, {Moses}, {Marley}, {Batalha}, {Kempton}, {Morley}, {Valenti}, \& {Vuitton}}]{He2023}
{He}, C., {Radke}, M., {Moran}, S.~E., {et~al.} 2023, arXiv e-prints, arXiv:2301.02745, \dodoi{10.48550/arXiv.2301.02745}

\bibitem[{{Holmberg} \& {Madhusudhan}(2023)}]{holmberg2023}
{Holmberg}, M., \& {Madhusudhan}, N. 2023, \mnras, 524, 377, \dodoi{10.1093/mnras/stad1580}

\bibitem[{{Hopkins} {et~al.}(2023){Hopkins}, {Archer}, {Bell}, {Suntharalingam}, \& {Todd}}]{Hopkins2023}
{Hopkins}, F.~E., {Archer}, S.~D., {Bell}, T.~G., {Suntharalingam}, P., \& {Todd}, J.~D. 2023, Nature Reviews Earth and Environment, 4, 361, \dodoi{10.1038/s43017-023-00428-7}

\bibitem[{Horne(1986)}]{horne_optimal_1986}
Horne, K. 1986, Publications of the Astronomical Society of the Pacific, 98, 609, \dodoi{10.1086/131801}

\bibitem[{{Hu} {et~al.}(2023){Hu}, {Gaillard}, \& {Kite}}]{Hu2023}
{Hu}, R., {Gaillard}, F., \& {Kite}, E.~S. 2023, \apjl, 948, L20, \dodoi{10.3847/2041-8213/acd0b4}

\bibitem[{{Hu} {et~al.}(2021)}]{Hu2021}
{Hu}, R., {et~al.} 2021, Astrophys. J. Lett., 921, L8, \dodoi{10.3847/2041-8213/ac1f92}

\bibitem[{Huang {et~al.}(2013)Huang, Freedman, Tashkun, Schwenke, \& Lee}]{HUANG2013}
Huang, X., Freedman, R.~S., Tashkun, S.~A., Schwenke, D.~W., \& Lee, T.~J. 2013, Journal of Quantitative Spectroscopy and Radiative Transfer, 130, 134, \dodoi{https://doi.org/10.1016/j.jqsrt.2013.05.018}

\bibitem[{Huang {et~al.}(2017)Huang, Schwenke, Freedman, \& Lee}]{huang2017}
Huang, X., Schwenke, D.~W., Freedman, R.~S., \& Lee, T.~J. 2017, Journal of Quantitative Spectroscopy and Radiative Transfer, 203, 224, \dodoi{https://doi.org/10.1016/j.jqsrt.2017.04.026}

\bibitem[{{Innes} {et~al.}(2023){Innes}, {Tsai}, \& {Pierrehumbert}}]{Innes2023}
{Innes}, H., {Tsai}, S.-M., \& {Pierrehumbert}, R.~T. 2023, arXiv e-prints, arXiv:2304.02698, \dodoi{10.48550/arXiv.2304.02698}

\bibitem[{{JWST Transiting Exoplanet Community Early Release Science Team} {et~al.}(2023){JWST Transiting Exoplanet Community Early Release Science Team}, {Ahrer}, {Alderson}, {Batalha}, {Batalha}, {Bean}, {Beatty}, {Bell}, {Benneke}, {Berta-Thompson}, {Carter}, {Crossfield}, {Espinoza}, {Feinstein}, {Fortney}, {Gibson}, {Goyal}, {Kempton}, {Kirk}, {Kreidberg}, {L{\'o}pez-Morales}, {Line}, {Lothringer}, {Moran}, {Mukherjee}, {Ohno}, {Parmentier}, {Piaulet}, {Rustamkulov}, {Schlawin}, {Sing}, {Stevenson}, {Wakeford}, {Allen}, {Birkmann}, {Brande}, {Crouzet}, {Cubillos}, {Damiano}, {D{\'e}sert}, {Gao}, {Harrington}, {Hu}, {Kendrew}, {Knutson}, {Lagage}, {Leconte}, {Lendl}, {MacDonald}, {May}, {Miguel}, {Molaverdikhani}, {Moses}, {Murray}, {Nehring}, {Nikolov}, {Petit dit de la Roche}, {Radica}, {Roy}, {Stassun}, {Taylor}, {Waalkes}, {Wachiraphan}, {Welbanks}, {Wheatley}, {Aggarwal}, {Alam}, {Banerjee}, {Barstow}, {Blecic}, {Casewell}, {Changeat}, {Chubb}, {Col{\'o}n}, {Coulombe}, {Daylan}, {de Val-Borro},
  {Decin}, {Dos Santos}, {Flagg}, {France}, {Fu}, {Garc{\'\i}a Mu{\~n}oz}, {Gizis}, {Glidden}, {Grant}, {Heng}, {Henning}, {Hong}, {Inglis}, {Iro}, {Kataria}, {Komacek}, {Krick}, {Lee}, {Lewis}, {Lillo-Box}, {Lustig-Yaeger}, {Mancini}, {Mandell}, {Mansfield}, {Marley}, {Mikal-Evans}, {Morello}, {Nixon}, {Ortiz Ceballos}, {Piette}, {Powell}, {Rackham}, {Ramos-Rosado}, {Rauscher}, {Redfield}, {Rogers}, {Roman}, {Roudier}, {Scarsdale}, {Shkolnik}, {Southworth}, {Spake}, {Steinrueck}, {Tan}, {Teske}, {Tremblin}, {Tsai}, {Tucker}, {Turner}, {Valenti}, {Venot}, {Waldmann}, {Wallack}, {Zhang}, \& {Zieba}}]{JWST_ERS2023}
{JWST Transiting Exoplanet Community Early Release Science Team}, {Ahrer}, E.-M., {Alderson}, L., {et~al.} 2023, \nat, 614, 649, \dodoi{10.1038/s41586-022-05269-w}

\bibitem[{{Karkoschka}(1998)}]{Karkoschka1998}
{Karkoschka}, E. 1998, \icarus, 133, 134, \dodoi{10.1006/icar.1998.5913}

\bibitem[{{Kasting} {et~al.}(1993){Kasting}, {Whitmire}, \& {Reynolds}}]{kasting1993}
{Kasting}, J.~F., {Whitmire}, D.~P., \& {Reynolds}, R.~T. 1993, Icarus, 101, 108, \dodoi{10.1006/icar.1993.1010}

\bibitem[{{Khare} {et~al.}(1984){Khare}, {Sagan}, {Arakawa}, {Suits}, {Callcott}, \& {Williams}}]{Khare1984}
{Khare}, B.~N., {Sagan}, C., {Arakawa}, E.~T., {et~al.} 1984, \icarus, 60, 127, \dodoi{10.1016/0019-1035(84)90142-8}

\bibitem[{{Kipping}(2013)}]{Kipping2013}
{Kipping}, D.~M. 2013, \mnras, 435, 2152, \dodoi{10.1093/mnras/stt1435}

\bibitem[{{Kite} \& {Barnett}(2020)}]{Kite_Barnett2020}
{Kite}, E.~S., \& {Barnett}, M.~N. 2020, Proceedings of the National Academy of Science, 117, 18264, \dodoi{10.1073/pnas.2006177117}

\bibitem[{{Kite} {et~al.}(2020)}]{Kite2020}
{Kite}, E.~S., {et~al.} 2020, Astrophys. J., 891, 111, \dodoi{10.3847/1538-4357/ab6ffb}

\bibitem[{{Knutson} {et~al.}(2014){Knutson}, {Benneke}, {Deming}, \& {Homeier}}]{Knutson2014}
{Knutson}, H.~A., {Benneke}, B., {Deming}, D., \& {Homeier}, D. 2014, \nat, 505, 66, \dodoi{10.1038/nature12887}

\bibitem[{Kochanov {et~al.}(2019)Kochanov, Gordon, Rothman, Shine, Sharpe, Johnson, Wallington, Harrison, Bernath, Birk, Wagner, Bris, Bravo, \& Hill}]{dms_cs2_1}
Kochanov, R., Gordon, I., Rothman, L., {et~al.} 2019, Journal of Quantitative Spectroscopy and Radiative Transfer, \dodoi{10.1016/j.jqsrt.2019.04.001}

\bibitem[{{Kubyshkina} {et~al.}(2018{\natexlab{a}}){Kubyshkina}, {Fossati}, {Erkaev}, {Johnstone}, {Cubillos}, {Kislyakova}, {Lammer}, {Lendl}, \& {Odert}}]{Kubyshkina2018a}
{Kubyshkina}, D., {Fossati}, L., {Erkaev}, N.~V., {et~al.} 2018{\natexlab{a}}, \aap, 619, A151, \dodoi{10.1051/0004-6361/201833737}

\bibitem[{{Kubyshkina} {et~al.}(2018{\natexlab{b}}){Kubyshkina}, {Fossati}, {Erkaev}, {Cubillos}, {Johnstone}, {Kislyakova}, {Lammer}, {Lendl}, \& {Odert}}]{Kubyshkina2018b}
---. 2018{\natexlab{b}}, \apjl, 866, L18, \dodoi{10.3847/2041-8213/aae586}

\bibitem[{{Leung} {et~al.}(2022)}]{Leung2022}
{Leung}, M., {et~al.} 2022, Astrophys. J., 938, 6, \dodoi{10.3847/1538-4357/ac8799}

\bibitem[{{Li} {et~al.}(2015){Li}, {Gordon}, {Rothman}, {Tan}, {Hu}, {Kassi}, {Campargue}, \& {Medvedev}}]{Li2015}
{Li}, G., {Gordon}, I.~E., {Rothman}, L.~S., {et~al.} 2015, \apjs, 216, 15, \dodoi{10.1088/0067-0049/216/1/15}

\bibitem[{{Line} {et~al.}(2011){Line}, {Vasisht}, {Chen}, {Angerhausen}, \& {Yung}}]{Line2011}
{Line}, M.~R., {Vasisht}, G., {Chen}, P., {Angerhausen}, D., \& {Yung}, Y.~L. 2011, \apj, 738, 32, \dodoi{10.1088/0004-637X/738/1/32}

\bibitem[{{Lodders} \& {Fegley}(2002)}]{Lodders2002}
{Lodders}, K., \& {Fegley}, B. 2002, Icarus, 155, 393, \dodoi{10.1006/icar.2001.6740}

\bibitem[{Lustig-Yaeger {et~al.}(2023)Lustig-Yaeger, Fu, May, Ceballos, Moran, Peacock, Stevenson, López-Morales, MacDonald, Mayorga, Sing, Sotzen, Valenti, Adams, Alam, Batalha, Bennett, Gonzalez-Quiles, Kirk, Kruse, Lothringer, Rustamkulov, \& Wakeford}]{lustig-yaeger_jwst_2023}
Lustig-Yaeger, J., Fu, G., May, E.~M., {et~al.} 2023, A {JWST} transmission spectrum of a nearby {Earth}-sized exoplanet,  arXiv.
\newblock \url{http://arxiv.org/abs/2301.04191}

\bibitem[{{Lustig-Yaeger} {et~al.}(2023){Lustig-Yaeger}, {Fu}, {May}, {Ortiz Ceballos}, {Moran}, {Peacock}, {Stevenson}, {L{\'o}pez-Morales}, {MacDonald}, {Mayorga}, {Sing}, {Sotzen}, {Valenti}, {Adams}, {Alam}, {Batalha}, {Bennett}, {Gonzalez-Quiles}, {Kirk}, {Kruse}, {Lothringer}, {Rustamkulov}, \& {Wakeford}}]{Lustig2023}
{Lustig-Yaeger}, J., {Fu}, G., {May}, E.~M., {et~al.} 2023, arXiv e-prints, arXiv:2301.04191, \dodoi{10.48550/arXiv.2301.04191}

\bibitem[{{MacDonald} \& {Madhusudhan}(2017)}]{macdonald2017}
{MacDonald}, R.~J., \& {Madhusudhan}, N. 2017, \mnras, 469, 1979, \dodoi{10.1093/mnras/stx804}

\bibitem[{Madhusudhan {et~al.}(2023)Madhusudhan, Moses, Rigby, \& Barrier}]{Madhusudhan2023}
Madhusudhan, N., Moses, J.~I., Rigby, F., \& Barrier, E. 2023, Faraday Discuss., 245, 80, \dodoi{10.1039/D3FD00075C}

\bibitem[{{Madhusudhan} {et~al.}(2021){Madhusudhan}, {Piette}, \& {Constantinou}}]{Madhusudhan2021}
{Madhusudhan}, N., {Piette}, A. A.~A., \& {Constantinou}, S. 2021, Astrophys. J., 918, 1, \dodoi{10.3847/1538-4357/abfd9c}

\bibitem[{Madhusudhan \& Seager(2009)}]{Madhusudhan2009}
Madhusudhan, N., \& Seager, S. 2009, \apj, 707, 24, \dodoi{10.1088/0004-637x/707/1/24}

\bibitem[{{Madhusudhan} \& {Seager}(2011)}]{madhu_seager2011}
{Madhusudhan}, N., \& {Seager}, S. 2011, \apj, 729, 41, \dodoi{10.1088/0004-637X/729/1/41}

\bibitem[{{Madhusudhan} {et~al.}(2016)}]{Madhu2016}
{Madhusudhan}, N., {et~al.} 2016, Space Sci. Rev., 205, 285, \dodoi{10.1007/s11214-016-0254-3}

\bibitem[{{Madhusudhan} {et~al.}(2020)}]{Madhusudhan2020}
---. 2020, Astrophys. J., 891, L7, \dodoi{10.3847/2041-8213/ab7229}

\bibitem[{{Meadows} {et~al.}(2022){Meadows}, {Graham}, {Abrahamsson}, {Adam}, {Amador-French}, {Arney}, {Barge}, {Barlow}, {Berea}, {Bose}, {Bower}, {Chan}, {Cleaves}, {Corpolongo}, {Currie}, {Domagal-Goldman}, {Dong}, {Eigenbrode}, {Enright}, {Fauchez}, {Fisk}, {Fricke}, {Fujii}, {Gangidine}, {Gezer}, {Glavin}, {Grenfell}, {Harman}, {Hatzenpichler}, {Hausrath}, {Henderson}, {Johnson}, {Jones}, {Hamilton}, {Hickman-Lewis}, {Jahnke}, {Kacar}, {Kopparapu}, {Kempes}, {Kish}, {Krissansen-Totton}, {Leavitt}, {Komatsu}, {Lichtenberg}, {Lindsay}, {Maggiori}, {Des Marais}, {Mathis}, {Morono}, {Neveu}, {Ni}, {Nixon}, {Olson}, {Parenteau}, {Perl}, {Quinn}, {Raj}, {Rodriguez}, {Rutter}, {Sandora}, {Schmidt}, {Schwieterman}, {Segura}, {Sekerci}, {Seyler}, {Smith}, {Soares}, {Som}, {Suzuki}, {Teece}, {Weber}, {Wolfe-Simon}, {Wong}, {Yano}, \& {Young}}]{Meadows2022}
{Meadows}, V., {Graham}, H., {Abrahamsson}, V., {et~al.} 2022, arXiv e-prints, arXiv:2210.14293, \dodoi{10.48550/arXiv.2210.14293}

\bibitem[{{Meadows} \& {Barnes}(2018)}]{Meadows2018}
{Meadows}, V.~S., \& {Barnes}, R.~K. 2018, in Handbook of Exoplanets, ed. H.~J. {Deeg} \& J.~A. {Belmonte}, 57, \dodoi{10.1007/978-3-319-55333-7_57}

\bibitem[{{Montet} {et~al.}(2015)}]{Montet2015}
{Montet}, B.~T., {et~al.} 2015, Astrophys. J., 809, 25, \dodoi{10.1088/0004-637X/809/1/25}

\bibitem[{Moran {et~al.}(2023)Moran, Stevenson, Sing, MacDonald, Kirk, Lustig-Yaeger, Peacock, Mayorga, Bennett, López-Morales, May, Rustamkulov, Valenti, Redai, Alam, Batalha, Fu, Gonzalez-Quiles, Highland, Kruse, Lothringer, Ceballos, Sotzen, \& Wakeford}]{moran_high_2023}
Moran, S.~E., Stevenson, K.~B., Sing, D.~K., {et~al.} 2023, High {Tide} or {Riptide} on the {Cosmic} {Shoreline}? {A} {Water}-{Rich} {Atmosphere} or {Stellar} {Contamination} for the {Warm} {Super}-{Earth} {GJ}{\textasciitilde}486b from {JWST} {Observations},  arXiv.
\newblock \url{http://arxiv.org/abs/2305.00868}

\bibitem[{{Moses} {et~al.}(2013)}]{Moses2013}
{Moses}, J.~I., {et~al.} 2013, Astrophys. J., 777, 34, \dodoi{10.1088/0004-637X/777/1/34}

\bibitem[{M\"{u}ller {et~al.}(2005)M\"{u}ller, Schl\"{o}der, Stutzki, \& Winnewisser}]{ocs_3}
M\"{u}ller, H., Schl\"{o}der, F., Stutzki, J., \& Winnewisser, G. 2005, Journal of Molecular Structure, 742, 215, \dodoi{10.1016/j.molstruc.2005.01.027}

\bibitem[{Nikitin {et~al.}(2016)Nikitin, Dmitrieva, \& Gordon}]{ch3cl_2}
Nikitin, A., Dmitrieva, T., \& Gordon, I. 2016, Journal of Quantitative Spectroscopy and Radiative Transfer, 177, 49, \dodoi{10.1016/j.jqsrt.2016.03.007}

\bibitem[{{Nixon} \& {Madhusudhan}(2021)}]{Nixon2021}
{Nixon}, M.~C., \& {Madhusudhan}, N. 2021, Mon. Notices Royal Astron. Soc., 505, 3414, \dodoi{10.1093/mnras/stab1500}

\bibitem[{{Orton} {et~al.}(2007)}]{orton2007}
{Orton}, G.~S., {et~al.} 2007, Icarus, 189, 544, \dodoi{10.1016/j.icarus.2007.02.003}

\bibitem[{{Phillips} {et~al.}(2022)}]{Phillips2022}
{Phillips}, C., {et~al.} 2022, arXiv e-prints, arXiv:2209.12919, \dodoi{10.48550/arXiv.2209.12919}

\bibitem[{{Phillips} {et~al.}(2021)}]{Phillips2021}
{Phillips}, C.~L., {et~al.} 2021, Astrophys. J., 923, 144, \dodoi{10.3847/1538-4357/ac29be}

\bibitem[{{Pierrehumbert}(2023)}]{Pierrehumbert2023}
{Pierrehumbert}, R.~T. 2023, \apj, 944, 20, \dodoi{10.3847/1538-4357/acafdf}

\bibitem[{{Piette} \& {Madhusudhan}(2020)}]{Piette2020}
{Piette}, A. A.~A., \& {Madhusudhan}, N. 2020, Astrophys. J., 904, 154, \dodoi{10.3847/1538-4357/abbfb1}

\bibitem[{{Pinhas} \& {Madhusudhan}(2017)}]{Pinhas2017}
{Pinhas}, A., \& {Madhusudhan}, N. 2017, \mnras, 471, 4355, \dodoi{10.1093/mnras/stx1849}

\bibitem[{{Pinhas} {et~al.}(2018)}]{Pinhas2018}
{Pinhas}, A., {et~al.} 2018, Mon. Notices Royal Astron. Soc., 480, 5314, \dodoi{10.1093/mnras/sty2209}

\bibitem[{{Polyansky} {et~al.}(2018){Polyansky}, {Kyuberis}, {Zobov}, {Tennyson}, {Yurchenko}, \& {Lodi}}]{Polyansky2018}
{Polyansky}, O.~L., {Kyuberis}, A.~A., {Zobov}, N.~F., {et~al.} 2018, \mnras, 480, 2597, \dodoi{10.1093/mnras/sty1877}

\bibitem[{{Pont} {et~al.}(2008){Pont}, {Knutson}, {Gilliland}, {Moutou}, \& {Charbonneau}}]{Pont2008}
{Pont}, F., {Knutson}, H., {Gilliland}, R.~L., {Moutou}, C., \& {Charbonneau}, D. 2008, \mnras, 385, 109, \dodoi{10.1111/j.1365-2966.2008.12852.x}

\bibitem[{{Radica} {et~al.}(2023){Radica}, {Welbanks}, {Espinoza}, {Taylor}, {Coulombe}, {Feinstein}, {Goyal}, {Scarsdale}, {Albert}, {Baghel}, {Bean}, {Blecic}, {Lafreni{\`e}re}, {MacDonald}, {Zamyatina}, {Allart1}, {Artigau}, {Batalha}, {Cook}, {Cowan}, {Dang}, {Doyon}, {Fournier-Tondreau}, {Johnstone}, {Line}, {Moran}, {Mukherjee}, {Pelletier}, {Roy}, {Talens}, {Filippazzo}, {Pontoppidan}, \& {Volk}}]{Radica2023}
{Radica}, M., {Welbanks}, L., {Espinoza}, N., {et~al.} 2023, \mnras, 524, 835, \dodoi{10.1093/mnras/stad1762}

\bibitem[{R\'{e}galia-Jarlot {et~al.}(2002)R\'{e}galia-Jarlot, Hamdouni, Thomas, der Heyden, \& Barbe}]{ocs_7}
R\'{e}galia-Jarlot, L., Hamdouni, A., Thomas, X., der Heyden, P.~V., \& Barbe, A. 2002, Journal of Quantitative Spectroscopy and Radiative Transfer, 74, 455, \dodoi{10.1016/S0022-4073(01)00267-9}

\bibitem[{{Richard} {et~al.}(2012){Richard}, {Gordon}, {Rothman}, {Abel}, {Frommhold}, {Gustafsson}, {Hartmann}, {Hermans}, {Lafferty}, {Orton}, {Smith}, \& {Tran}}]{richard2012}
{Richard}, C., {Gordon}, I.~E., {Rothman}, L.~S., {et~al.} 2012, \jqsrt, 113, 1276, \dodoi{10.1016/j.jqsrt.2011.11.004}

\bibitem[{{Rothman} {et~al.}(2010){Rothman}, {Gordon}, {Barber}, {Dothe}, {Gamache}, {Goldman}, {Perevalov}, {Tashkun}, \& {Tennyson}}]{rothman2010}
{Rothman}, L.~S., {Gordon}, I.~E., {Barber}, R.~J., {et~al.} 2010, \jqsrt, 111, 2139, \dodoi{10.1016/j.jqsrt.2010.05.001}

\bibitem[{{Rustamkulov} {et~al.}(2023){Rustamkulov}, {Sing}, {Mukherjee}, {May}, {Kirk}, {Schlawin}, {Line}, {Piaulet}, {Carter}, {Batalha}, {Goyal}, {L{\'o}pez-Morales}, {Lothringer}, {MacDonald}, {Moran}, {Stevenson}, {Wakeford}, {Espinoza}, {Bean}, {Batalha}, {Benneke}, {Berta-Thompson}, {Crossfield}, {Gao}, {Kreidberg}, {Powell}, {Cubillos}, {Gibson}, {Leconte}, {Molaverdikhani}, {Nikolov}, {Parmentier}, {Roy}, {Taylor}, {Turner}, {Wheatley}, {Aggarwal}, {Ahrer}, {Alam}, {Alderson}, {Allen}, {Banerjee}, {Barat}, {Barrado}, {Barstow}, {Bell}, {Blecic}, {Brande}, {Casewell}, {Changeat}, {Chubb}, {Crouzet}, {Daylan}, {Decin}, {D{\'e}sert}, {Mikal-Evans}, {Feinstein}, {Flagg}, {Fortney}, {Harrington}, {Heng}, {Hong}, {Hu}, {Iro}, {Kataria}, {Kempton}, {Krick}, {Lendl}, {Lillo-Box}, {Louca}, {Lustig-Yaeger}, {Mancini}, {Mansfield}, {Mayne}, {Miguel}, {Morello}, {Ohno}, {Palle}, {Petit dit de la Roche}, {Rackham}, {Radica}, {Ramos-Rosado}, {Redfield}, {Rogers}, {Shkolnik}, {Southworth}, {Teske}, {Tremblin},
  {Tucker}, {Venot}, {Waalkes}, {Welbanks}, {Zhang}, \& {Zieba}}]{Rustamkulov2023}
{Rustamkulov}, Z., {Sing}, D.~K., {Mukherjee}, S., {et~al.} 2023, \nat, 614, 659, \dodoi{10.1038/s41586-022-05677-y}

\bibitem[{{Scheucher} {et~al.}(2020){Scheucher}, {Wunderlich}, {Grenfell}, {Godolt}, {Schreier}, {Kappel}, {Haus}, {Herbst}, \& {Rauer}}]{Scheucher2020}
{Scheucher}, M., {Wunderlich}, F., {Grenfell}, J.~L., {et~al.} 2020, \apj, 898, 44, \dodoi{10.3847/1538-4357/ab9084}

\bibitem[{{Schlichting} \& {Young}(2022)}]{Schlichting2022}
{Schlichting}, H.~E., \& {Young}, E.~D. 2022, \psj, 3, 127, \dodoi{10.3847/PSJ/ac68e6}

\bibitem[{{Schwieterman} {et~al.}(2018){Schwieterman}, {Kiang}, {Parenteau}, {Harman}, {DasSarma}, {Fisher}, {Arney}, {Hartnett}, {Reinhard}, {Olson}, {Meadows}, {Cockell}, {Walker}, {Grenfell}, {Hegde}, {Rugheimer}, {Hu}, \& {Lyons}}]{schwieterman2018}
{Schwieterman}, E.~W., {Kiang}, N.~Y., {Parenteau}, M.~N., {et~al.} 2018, Astrobiology, 18, 663, \dodoi{10.1089/ast.2017.1729}

\bibitem[{{Seager} {et~al.}(2013{\natexlab{a}}){Seager}, {Bains}, \& {Hu}}]{seager2013a}
{Seager}, S., {Bains}, W., \& {Hu}, R. 2013{\natexlab{a}}, \apj, 775, 104, \dodoi{10.1088/0004-637X/775/2/104}

\bibitem[{{Seager} {et~al.}(2013{\natexlab{b}}){Seager}, {Bains}, \& {Hu}}]{seager2013b}
---. 2013{\natexlab{b}}, \apj, 777, 95, \dodoi{10.1088/0004-637X/777/2/95}

\bibitem[{{Seager} {et~al.}(2016){Seager}, {Bains}, \& {Petkowski}}]{seager2016}
{Seager}, S., {Bains}, W., \& {Petkowski}, J.~J. 2016, Astrobiology, 16, 465, \dodoi{10.1089/ast.2015.1404}

\bibitem[{Seager {et~al.}(2013)}]{Seager2013}
Seager, S., {et~al.} 2013, Astrophys. J., 777, 95

\bibitem[{{Segura} {et~al.}(2005){Segura}, {Kasting}, {Meadows}, {Cohen}, {Scalo}, {Crisp}, {Butler}, \& {Tinetti}}]{Segura2005}
{Segura}, A., {Kasting}, J.~F., {Meadows}, V., {et~al.} 2005, Astrobiology, 5, 706, \dodoi{10.1089/ast.2005.5.706}

\bibitem[{Sharpe {et~al.}(2004)Sharpe, Johnson, Sams, Chu, Rhoderick, \& Johnson}]{dms_cs2_2}
Sharpe, S.~W., Johnson, T.~J., Sams, R.~L., {et~al.} 2004, Applied Spectroscopy, 58, 1452, \dodoi{10.1366/0003702042641281}

\bibitem[{{Skilling}(2004)}]{Skilling2004}
{Skilling}, J. 2004, in American Institute of Physics Conference Series, Vol. 735, Bayesian Inference and Maximum Entropy Methods in Science and Engineering: 24th International Workshop on Bayesian Inference and Maximum Entropy Methods in Science and Engineering, ed. R.~{Fischer}, R.~{Preuss}, \& U.~V. {Toussaint}, 395--405, \dodoi{10.1063/1.1835238}

\bibitem[{{Stevenson} {et~al.}(2010){Stevenson}, {Harrington}, {Nymeyer}, {Madhusudhan}, {Seager}, {Bowman}, {Hardy}, {Deming}, {Rauscher}, \& {Lust}}]{Stevenson2010}
{Stevenson}, K.~B., {Harrington}, J., {Nymeyer}, S., {et~al.} 2010, \nat, 464, 1161, \dodoi{10.1038/nature09013}

\bibitem[{Sung {et~al.}(2009)Sung, Toth, Brown, \& Crawford}]{ocs_5}
Sung, K., Toth, R., Brown, L., \& Crawford, T. 2009, Journal of Quantitative Spectroscopy and Radiative Transfer, 110, 2082, \dodoi{10.1016/j.jqsrt.2009.05.013}

\bibitem[{{Tian} \& {Heng}(2023)}]{Tian2023}
{Tian}, M., \& {Heng}, K. 2023, arXiv e-prints, arXiv:2301.10217, \dodoi{10.48550/arXiv.2301.10217}

\bibitem[{Toth {et~al.}(2010)Toth, Sung, Brown, \& Crawford}]{ocs_6}
Toth, R., Sung, K., Brown, L., \& Crawford, T. 2010, Journal of Quantitative Spectroscopy and Radiative Transfer, 111, 1193, \dodoi{10.1016/j.jqsrt.2009.10.014}

\bibitem[{{Trotta}(2008)}]{Trotta2008}
{Trotta}, R. 2008, Contemporary Physics, 49, 71, \dodoi{10.1080/00107510802066753}

\bibitem[{{Tsai} {et~al.}(2021)}]{Tsai2021}
{Tsai}, S.-M., {et~al.} 2021, Astrophys. J. Lett., 922, L27, \dodoi{10.3847/2041-8213/ac399a}

\bibitem[{{Tsiaras} {et~al.}(2019)}]{Tsiaras2019}
{Tsiaras}, A., {et~al.} 2019, Nat. Astron., 3, 1086, \dodoi{10.1038/s41550-019-0878-9}

\bibitem[{{Welbanks} {et~al.}(2019){Welbanks}, {Madhusudhan}, {Allard}, {Hubeny}, {Spiegelman}, \& {Leininger}}]{Welbanks2019}
{Welbanks}, L., {Madhusudhan}, N., {Allard}, N.~F., {et~al.} 2019, \apjl, 887, L20, \dodoi{10.3847/2041-8213/ab5a89}

\bibitem[{{Yu} {et~al.}(2021)}]{Yu2021}
{Yu}, X., {et~al.} 2021, Astrophys. J., 914, 38, \dodoi{10.3847/1538-4357/abfdc7}

\bibitem[{Yung(1999)}]{Yung1999}
Yung, Y. L. Y.~L. 1999, Photochemistry of planetary atmospheres / Yuk L. Yung, William B. DeMore., Oxford scholarship online

\bibitem[{{Zhang} {et~al.}(2005){Zhang}, {Sun}, {Zhou}, \& {Wang}}]{Zhang2005}
{Zhang}, Q., {Sun}, T., {Zhou}, X., \& {Wang}, W. 2005, Chemical Physics Letters, 414, 316, \dodoi{10.1016/j.cplett.2005.08.084}

\end{thebibliography}
\bibliographystyle{aasjournal}

\end{document}